\documentclass[journal,twocolumn]{IEEEtran}
\usepackage{cite}
\usepackage{amsmath,amsfonts,amsxtra,amssymb,latexsym,amscd,amsthm,mathrsfs,bm}
\usepackage{mathtools}
\usepackage{tabularx}
\usepackage{graphicx}
\usepackage{hyperref}
\usepackage{xcolor}
\usepackage{algorithm}
\usepackage{algorithmic}
\usepackage{cite}
\usepackage[colorinlistoftodos,textsize=tiny]{todonotes}
\usepackage{balance}
\usepackage{epstopdf}
\usepackage{balance}
\usepackage{booktabs}
\usepackage{lipsum}

\usepackage{caption}
\usepackage{subcaption}
\newcommand{\bs}{\boldsymbol}
\newcommand{\nn}{\nonumber}

\usepackage{eqparbox}

\newtheorem{lemma}{Lemma}
\newtheorem{corollary}{Corollary}

\usepackage{etoolbox}  % patch def of algorithmic environment
\makeatletter
\patchcmd{\algorithmic}{\addtolength{\ALC@tlm}{\leftmargin} }{\addtolength{\ALC@tlm}{\leftmargin}}{}{}
\makeatother

\newcommand{\T}{\mathsf{T}}

\newcommand{\Hi}{\mathsf{H}}

\title{\LARGE Novel Double-Chirp Preamble Design  for Multiuser Asynchronous  Massive MIMO LoRa Networks
	\thanks{The authors are with the Department of Electrical and Computer Engineering, University of Saskatchewan, Saskatoon, Canada S7N5A9. Emails: \{khai.nguyen,  e.bedeer\}@usask.ca.}
}

\author{\IEEEauthorblockN{The Khai Nguyen and Ebrahim Bedeer}}

\begin{document}
	\maketitle	
	\vspace{-10cm}
	
	\begin{abstract}
		This paper proposes a novel preamble design and detection method for multiuser asynchronous massive MIMO LoRa networks. Unlike existing works, which only consider the preamble detection for a single  end device (ED), we propose to simultaneously detect the preambles of multiple EDs that asynchronously transmit their uplink (UL) packets to a multiple-antenna gateway (GW). First, we show that the preamble detection in multiuser LoRa networks with the conventional  \emph{single-chirp} { preamble} suffers  from the so-called \emph{preamble resemblance} effect. This means that the preamble of any single ED can resemble the preambles of all EDs in the network and make it impossible to determine to which ED a preamble belongs.  To address this problem, a novel \emph{double-chirp} preamble design and a preamble assignment method are { proposed}, which can mitigate the\emph{ preamble resemblance} effect by making the preamble of each ED unique and recognizable. Next, a maximum-likelihood (ML) based detection scheme for the proposed \emph{double-chirp} preamble is derived. Finally, since the proposed algorithm requires the calculation of the discrete Fourier transform (DFT) every sampling period, we propose a low-complexity technique to calculate the DFT recursively to reduce the complexity of our proposed design. Simulation shows that the proposed preamble  design and detection  require just about 2 dB more power  to increase the number of EDs from one to 15 in the Rayleigh fading channel while achieving the same preamble detection error performance.
	\end{abstract}
	\vspace*{-0.2cm}
	\begin{IEEEkeywords}
		LoRaWAN, massive MIMO, multiuser preamble detection.
	\end{IEEEkeywords}
	\vspace*{-0.5cm}
	\section{Introduction}\label{Sec:Intro}
	LoRaWAN is a  standard for low-power wide-area networks that has been well received as a key technology   for Internet of Things (IoT) \cite{Milarokostas2023,Maleki2024,Hu2022,Zhang2024,Tao2025}. In the physical layer, LoRa adopts the chirp spread spectrum (CSS) modulation, also known as the LoRa modulation. The CSS signal has  linearly-varying frequency and  constant signal envelopes, which can provide the robustness against Doppler shift, and can offer very long-range, low-power communication, { yet can be implemented efficiently with the simple discrete Fourier transform (DFT) based receiver \cite{Azim2023}}. In the medium access control (MAC) layer, LoRa uses pure ALOHA protocol, which, despite having low complexity, has a high risk of packet collision when end devices (EDs) operate with the same spreading factors (SFs) \cite{Mu2026}. Thus, LoRa often resorts to a duty cycle limit to avoid collisions. This results in a bottleneck in both the capacity and connectivity of LoRaWAN in environments with high ED density, which are the current and future scenarios of modern IoT networks. Solving this problem to enable concurrent  transmission-reception in LoRaWAN requires the solutions of two subproblems: preamble detection and  multiuser detection.
	
	In terms of multiuser detection, to deal with packet collisions in LoRaWAN,  { the works in} \cite{Wang2019, Tapparel2021, AlHamdani2021,Petroni2022} utilize successive interference cancellation (SIC) to detect collided packets from different EDs. The payload of the EDs is decoded in a successive manner, and then subtracted before detecting the payloads of the succeeding packets.  In \cite{Xhonneux2022}, the time-frequency offsets and power characteristics of different LoRa packets are exploited to develop a maximum likelihood { (ML)} detector for the case of two EDs. In \cite{Nguyen2023a}, a non-coherent detector for multiuser GW massive MIMO LoRa systems is studied. The cooperative GWs differentiate EDs based on their distinctive relative locations from the GWs and the unique received power signatures. In \cite{Jha2022,Nguyen2025}, multiuser LoRa reception  is realized with coherent detection, which takes advantage of the asymptotic orthogonality among channels of different EDs. In \cite{Ortin2019}, the authors proposed  listen-before-talk  protocols  to reduce collision in LoRaWAN. The scheme is developed for  both the physical and MAC layer, which demonstrates collision reduction performance by sacrificing complexity and transmit power. 
	
	%Most of the aforementioned works \cite{Wang2019, Tapparel2021, AlHamdani2021, Xhonneux2022, Jha2022}  requires channel state information (CSI), which means extra overhead for pilot training.
	%However, channel estimation in pure ALOHA LoRa is challenging due to \textit{asynchronous} transmission, which can cause interference between different EDs' training sequences and payloads. In addition, the overhead associated with channel estimation, i.e., sending pilot symbols, is resource-expensive in LoRa networks due to the low data rate. 
	%	To address this overhead, in \cite{Nguyen2021,Nguyen2023}, the authors  propose a semi-coherent detection method. The channel is estimated iteratively based on the detected data, which does not require any overhead associated with pilot training. In \cite{Jha2022,Nguyen2025,Kang2024}, channel estimation methods based on preambles for multiuser LoRa networks are presented. {However, one critical requirement for the methods in \cite{Jha2022,Kang2024} is that signals from different EDs must be \textit{synchronized} at packet level (i.e., preambles from all EDs must be aligned),  which indicates that  slotted ALOHA is required}. Whereas, { the method in} \cite{Nguyen2025} can work even with pure ALOHA with asynchronous packets.
	
	In order to realize multiuser detection, preamble detection is the first step and hence, is  of highest importance. However, this topic has not received adequate attention in the literature, and mostly considers single-ED preamble detection. The majority of existing works in LoRa preamble detection is based on \emph{majority vote} \cite{Ghanaatian2019,Tang2019,Edward2019,bernier2020}. First, the  signal is processed with the conventional discrete Fourier transform (DFT)-based receiver. Then, the power of the  signal in the frequency domain is compared with a preset threshold to detect the presence of the preamble. If the threshold is surpassed at the same  frequency bin multiple times over multiple consecutive DFT blocks, the GW  determines that a preamble is present. The condition for the number of power peaks that surpass the threshold to indicate  a preamble can vary. Experimentally, at least 4 consecutive peaks on the same frequency bin must surpass the threshold to declare the presence of a preamble \cite{Edward2019}. In \cite{Edward2019}, the authors propose to relax this condition such that the peaks are not required to be consecutive, but must still remain within the preamble length. The relaxed condition results in a lower probability that the GW misses the presence of a preamble. In\cite{Ghanaatian2019}, the authors propose to count the peaks that surpass the threshold on multiple adjacent bins rather than just one single bin as conventional. This modification accounts for the Doppler effect that causes undesirable frequency offset to the bins in the DFT domain, and results in an improved performance. In \cite{Tang2019,bernier2020}, the peaks are obtained by averaging the power of multiple consecutive DFT blocks to obtain a better performance at low SNR.
	However, both the  threshold  and the number of peaks are not optimized, but rather chosen arbitrarily, which leads to  suboptimal performance \cite{Kang2022}. 
	
	To improve the preamble detection performance, in \cite{Kang2022,Tapparel2024}, the authors proposed  preamble detection methods that consider the effect of noise and interference to find  optimal  thresholds for preamble detection. Unlike previous works that only focus on the aspect of maximizing the preamble detection rate, the work in \cite{Kang2022} also takes into account the effect of false alarm, which can have a much more negative impact on the performance of LoRaWAN. In \cite{Kang2023}, the author extended the work in \cite{Kang2022} for fading channels with both coherent and non-coherent detection. A closed-form expression for the threshold was derived, which leads to a significantly better performance compared to  previous  works. In \cite{Wang2024}, a detailed study of LoRa preamble detection based on the hypothesis model in \cite{Kang2022} is performed under the consideration of timing and frequency offset, which results in a more robust and complete design. In \cite{Vangelista2022}, the design for LoRa synchronization and start-of-frame detection is investigated. The properties of the transmitted and received preamble signal are studied, which results in an algorithm that can simultaneously perform carrier frequency-offset, timing-offset, and start-of-frame recovery.
	
	Although some of the aforementioned works did take the effect of \emph{inter-user interference} into account, only the preamble detection for a single target ED is considered. Moreover, the interference caused by data payloads and preambles are generalized as a common  \emph{inter-user interference} and thus, oversimplified. To the best of our knowledge, the effect of interfering preambles to the preamble of a different ED has not been sufficiently investigated. In LoRaWAN, where multiple EDs concurrently transmit on the same time-frequency resources, using the same spreading factor (SF),  even if a preamble is successfully detected, it is impossible to determine to which ED that preamble belongs. In this paper, we will show that the culprit is the so-called \emph{preamble resemblance} effect. This means that a conventional LoRa preamble of any ED can resemble the preambles of any other EDs. Consequently, the GW cannot determine the identity of the detected ED only based on the information in the preamble.  
	
	Therefore, in this paper, we propose a novel preamble design that  enables the multiuser preamble detection, with the aid of a massive MIMO GW.
	The proposed design uses  \emph{double-chirp} preambles, which means that a preamble is the sum of two different chirps, that are uniquely assigned to only one ED in the network. Although multi-chirp preambles have been considered in literature, the usage is only limited to channel estimation \cite{Kang2024}, or collision survival for one packet only \cite{Jiang2024}, while the preamble detection of multiple overlapped packets remains a research gap. On the other hand, in this paper, double-chirp preambles  are designed and assigned in a way such that preamble resemblance can be avoided, which enables preamble detection for multiple EDs simultaneously. Moreover, the massive MIMO GW provides the distinction among the channels of different EDs to the GW \cite{Marzetta2016}, which  provides the robustness against noise and interference in preamble detection.	
	An ML-based { non-coherent} preamble  detection algorithm is then derived to simultaneously detect the preambles of multiple EDs. Next, since the proposed algorithm requires the calculation of the DFT every sampling period, we proposed a recursive technique to replace the fast Fourier transform (FFT) used in the DFT-based LoRa receiver. The proposed technique can reduce the complexity of calculating the DFT in our proposed technique by order of the $\mathrm{SF}$. The contributions of the paper are summarized as follows:
		\begin{itemize}
			\item A novel double-chirp preamble design for multiuser asynchronous MIMO LoRa networks is proposed to  detect preambles from multiple EDs simultaneously, which mitigates packet-loss due to  collision.
			\item The preamble resemblance effect that degrades preamble detection performance is analyzed, and addressed with proper design and assignment of the proposed double-chirp preambles.
			\item An ML-based non-coherent preamble detection scheme is derived to simultaneously detect the double-chirp preambles of multiple EDs in the network.
			\item A technique to reduce the complexity of the proposed design is introduced, which calculates the DFT used in the receiver in a recursive manner. This eliminates the need to perform FFT every sample period and reduces the complexity by the order of the SF.
		\end{itemize}

	The remainder of this paper is organized as follows. Section~\ref{Sec:sys} presents the multiuser LoRa system model and reviews fundamental background on LoRa modulation, demodulation and preamble detection. Section~\ref{Sec:preamble} introduces the novel double-chirp preamble design and analyzes the \emph{preamble resemblance} effect. Section~\ref{Sec:ML} derives the ML-based preamble detector. { Section \ref{Sec:com} analyzes the complexity of the proposed double-chirp preamble ML-based detection, and proposes a complexity reduction method for the proposed design.} Finally, Section~\ref{Sec:sim} provides simulation results, and Section~\ref{Sec:con} concludes the paper.

	\textit{Notations:} The following notations are used throughout the paper. Transpose matrices are denoted by $[\cdot]^{\T}$. Hermitian matrices are denoted by $[\cdot]^{\Hi}$.  The averaging (expectation) of a random variable or random vector is denoted by $\mathbb{E}\left\{\cdot\right\}$. The norm of a vector is denoted by $\Vert\cdot\Vert$. The real and imaginary parts of a complex number are denoted by $\mathfrak{R}\{\cdot\}$ and $\mathfrak{I}\{\cdot\}$, respectively. The modulo operator is denoted by $\mathrm{mod}$. The set of $d_1\times d_2$ complex matrices is denoted by $\mathbb{C}^{d_1\times d_2}$. The complex normal distribution with variance $\sigma^2$ is denoted by $\mathcal{CN}\left(0,\sigma^2\right)$. The complex conjugate is denoted by $(\cdot)^{*}$. The determinant of a squared matrix is $\mathrm{det}(\cdot)$. The pairwise multiplication is denoted by $\otimes$. { Circular shift operations of a vector $\boldsymbol{a}$ by $b$ elements to the left and the right are denoted by $\boldsymbol{a}\ll b$ and $\boldsymbol{a}\gg b$, respectively.}

	\section{Multiuser LoRaWAN System Model and Challenges with Conventional Preamble Design}\label{Sec:sys}
	This section introduces the multiuser LoRa system model and reviews the fundamentals of LoRa detection with the DFT-based receiver, conventional LoRa preamble structure and its challenges when being implemented in a multiuser scenario.
	\subsection{System model}
\begin{figure}[t!]
		\centering
		\includegraphics[width=0.5\textwidth]{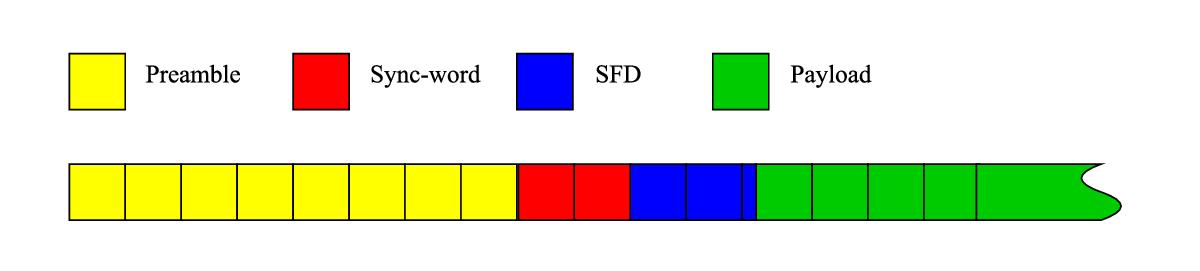}
		\caption{LoRa packet structure.}
		%\label{fig.1}
		\label{fig-LoRa-frame}
	\end{figure}

	Consider a LoRa network in which $N_u$ single-antenna EDs simultaneously transmit uplink (UL)  packets to a GW  with $L$ antennas. With pure ALOHA, the packets from different EDs are asynchronous.  Each LoRa packet has multiple symbols, which are drawn from a  set of $M=2^{\mathrm{SF}}$ linearly-frequency-variant LoRa symbols (also known as  chirps).  Specifically, the $\tilde{m}$th chirp is defined as $\bs{s}_{\tilde{m}}=[s_{\tilde{m},1},s_{\tilde{m},2},\ldots,s_{\tilde{m},M}]^{\T}$, where $\tilde{m}=0,1,\ldots,M-1$. The $0$th chirp is known as the basic upchirp, whose samples are defined as:
	\begin{equation}
		s_{0,m}=\exp\left\{j2\pi\left(\frac{ (m-1)^2}{2M}-\frac{m-1}{2}\right)\right\}.
	\end{equation}
	Then, the $m$th sample of the $\tilde{m}$th chirp can be obtained from $s_{0,m}$ as $s_{\tilde{m},m}=s_{0,m+\tilde{m}}$. From this chirp set, a LoRa packet is constructed with the structure  in Fig.~\ref{fig-LoRa-frame}, which starts with $N$ consecutive identical basic upchirps $\boldsymbol{s}_0$ called the \emph{preamble}, followed by 2 symbols known as \emph{sync-words} (used for network identifier), then 2.25 basic downchirp $\boldsymbol{s}^*_0$ of \emph{start-frame delimiter (SFD)} marking the beginning of the data.
	
	Since each LoRa symbol consists of $M$ discrete samples, the detection of each LoRa symbol must be performed collectively on $M$ consecutive samples in the time domain.
	Let
	$\mathbf{Y}_n=[\boldsymbol{y}_{n,1},\boldsymbol{y}_{n,2}\dots \boldsymbol{y}_{n,L}]^{\T}\in\mathbb{C}^{L\times M}$ be the signal received on $L$ antennas at the GW during one LoRa symbol period from the discrete time $n$ to $n+M-1$, where $
	\boldsymbol{y}_{n,\ell} = [y_{n,\ell},y_{n+1,\ell}\dots y_{n+M-1,\ell}]^{\T}\in\mathbb{C}^{M\times1},
	$
	%%%%%%%%%%%%%%%%
	is the received signal on the $\ell$th antennas  from the discrete time $n$ to $n+M-1$.
	Consider a Rayleigh fading channel, $\mathbf{Y}_n$ can be expressed  as:
	%%%%%%%%%%%
	\begin{equation}\label{eq.data}
		\small
		\mathbf{Y}_n =\sum_{u=1}^{N_u}\boldsymbol{h}_u\bar{\boldsymbol{x}}_{n,u}^{\T} + \boldsymbol{\Omega}_n,
	\end{equation}
	%%%%%%%%%%%%%%%
	where  $\boldsymbol{h}_u\sim\mathcal{CN}(0,\textbf{I}_{L})$ denotes the uncorrelated Rayleigh fading channel from the $u$th ED to the GW  ($\textbf{I}_{L}$ is the identity matrix of size $L$), $\bar{\boldsymbol{x}}_{n,u}\in\mathbb{C}^{M\times 1}$ is the $M$-sample signal vector of the $u$th ED during one LoRa symbol period, and $\boldsymbol{\Omega}_n\in\mathbb{C}^{L\times M}$ is the AWGN noise matrix with i.i.d. elements ${\omega}_{\ell,m}\sim\mathcal{CN}(0,\sigma^2)$.
	%Similarly to when estimating the channel of the ${\prm}$th ED with its preamble, 
	Note that  the signal from each ED $\bar{\boldsymbol{x}}_{n,u}\in\mathbb{C}^{M\times 1}$ can be partially or completely preamble and/or payload. This is  due to the asynchronous nature of ALOHA at MAC layer,   where packets transmitted from different EDs are not necessarily synchronized.

		\textbf{Remark:} The assumption of  uncorrelated Rayleigh fading channels is well justified for LoRa networks employing multiple antennas at the GW, particularly as LoRa technology evolves toward higher frequency bands. While traditional LoRa deployments operate in sub-GHz bands, Semtech has introduced commercial LoRa products in the 2.4 GHz  band \cite{Semtech2}. At this frequency, the wavelength is significantly shorter than in sub-GHz systems. Consequently, practical GW deployments can accommodate multiple antennas with sufficient separation to experience nearly independent fading channels. In rich-scattering environments where a dominant line-of-sight component is absent, the channel coefficients can therefore be reasonably modeled as independent Rayleigh random variables. This assumption simplifies analytical performance evaluation while accurately reflecting the propagation characteristics of multi-antenna LoRa GWs operating in emerging GHz-band.
	%	\begin{figure}[t!]
		%		\centering
		%		\includegraphics[width=0.5\textwidth]{LoRaframe2_v4.pdf}
		%		\caption{Signal model for the $\prm$th ED.}
		%		%\label{fig.1}
		%		\label{fig-LoRa-frameq}
		%	\end{figure}
	%\vspace{length}
	\subsection{LoRa data and preamble detection in single-ED scenario}
	The most common LoRa modulation method is the  DFT-based receiver, which can be performed  on the signal $\boldsymbol{y}_{n,\ell}$  separately on each and every antenna. First,   $\boldsymbol{y}_{n,\ell}$ is dechirped with the conjugate of the basic chirp $\boldsymbol{s}_0$, then transformed with the DFT to obtain the power distribution on $M$ frequency bins as ${\boldsymbol{V}}_{n,\ell}=[{{V}}_{n,\ell}[0],{{V}}_{n,\ell}[1]\dots {{V}}_{n,\ell}[M-1]]^{\T}$:
	%%%%%%%%%%%%%%%%%%%
	\begin{equation}\label{eq-pdw-V}
		{\boldsymbol{V}}_{n,\ell}= \mathrm{DFT}\left\{\boldsymbol{y}_{n,\ell}\otimes\boldsymbol{s}_0^*\right\}\in \mathbb{C}^{M\times1}.
	\end{equation}
	%%%%%%%%%%%%%%%%%%
	%where $\otimes$ denotes element-wise multiplication. The DFT  output $	{\boldsymbol{V}}_{n,\ell}$ is dependent on the synchronization of the symbol.  
	
\begin{figure}[t!]
	\centering
	\includegraphics[width=0.5\textwidth]{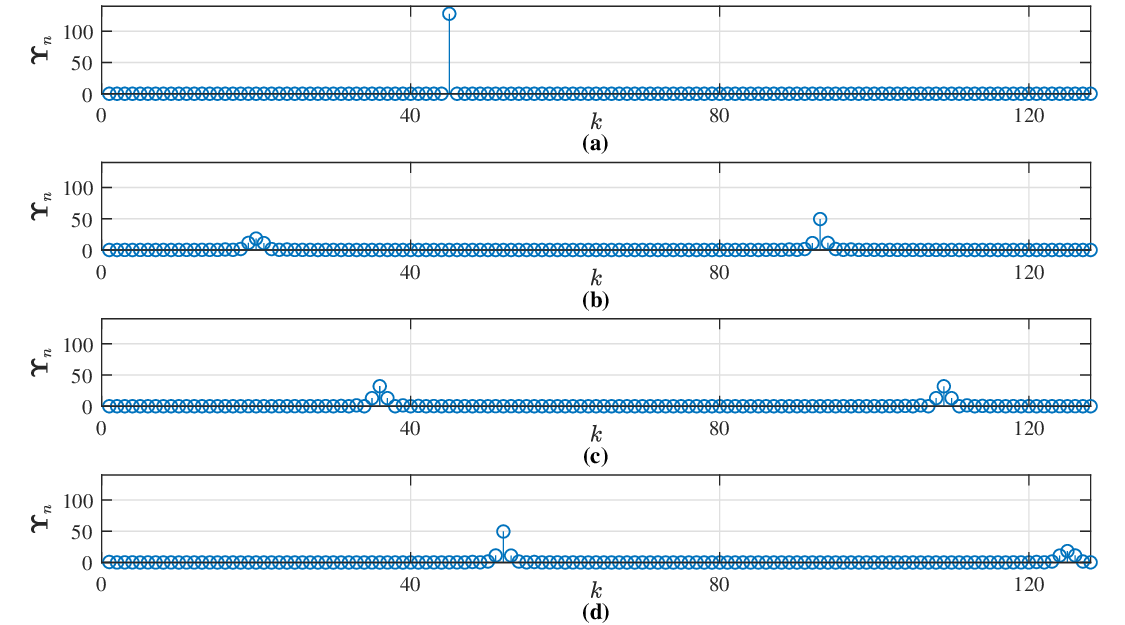}
	\caption{DFT output of the dechirped windown with $\boldsymbol{s}_{44}$ and $\boldsymbol{s}_{99}$ ($\mathrm{SF}=7$). (a) $\tau = 0$, (b) $\tau = 48$, (c) $\tau = 64$, (d) $\tau = 80$.}
	%\label{fig.1}
	\label{fig-bin}
\end{figure}

	Suppose only the $u$th ED is transmitting, if the signal $\boldsymbol{y}_{n,\ell}$ received on the $\ell$th antenna is perfectly synchronized with a transmitted LoRa chirp, say $\boldsymbol{s}_{\tilde{m}}$, all of  the chirp power  will be focused on the ${\tilde{m}}$th  bin, while the remaining bins receive zero power from that chirp\cite{Nguyen2021}. Mathematically, suppose:
	\begin{equation}
		\boldsymbol{y}_{n,\ell} = {h}_{u,\ell}{\boldsymbol{s}}_{\tilde{m}} + \boldsymbol{\omega}_{n,\ell},
	\end{equation}
	%%%%%%%%%%%%%%%%%%
	where $h_{u,\ell}$ is the channel from the $u$th ED to the GW on the $\ell$th antenna (the $\ell$th element of $\boldsymbol{h}_{u}$), and $\boldsymbol{\omega}_{n,\ell}$ is the AWGN noise sample vector on the $\ell$th antennas from time $n$ to $n+M-1$ (the $\ell$th row of $\boldsymbol{\Omega}_n^{\T}$). When performing the dechirping, the dechirped LoRa chirp component ${\boldsymbol{s}}_{\tilde{m}}\otimes{\boldsymbol{s}}_{0}^{*}$ becomes:
	%%%%%%%%%%%%%%%%
	%\lipsum[1][1-4]
	\begin{align}\label{eq7}
		%\begin{split}
		&s_{\tilde{m},m}s_{0,m}^*=\exp\left\{j2\pi\left(\frac{ (m+\tilde{m}-1)^2}{2M}-\frac{m+\tilde{m}-1}{2}\right)\right\}\nn\\
		&\times\exp\left\{-j2\pi\left(\frac{ (m-1)^2}{2M}-\frac{m-1}{2}\right)\right\}\nn\\
		&=\underbrace{\exp\left\{j2\pi\left(\frac{ \tilde{m}^2}{2M}-\frac{\tilde{m}}{2}\right)\right\}}_{\text{constant phase } \theta_{\tilde{m}}}\underbrace{\exp\left\{\frac{j2\pi \tilde{m}(m-1)}{M}\right\}}_{\text{linear phase}}.
		%	\end{split}
\end{align}
%%%%%%%%%%%%%%%%%%
As can be seen, the quadratic phase component of the LoRa chirp  ${\boldsymbol{s}}_{\tilde{m}}$ has been eliminated by  dechirping, and only the linear phase component remains with the discrete frequency of $\tilde{m}$. Consequently, as can be seen in Fig.~\ref{fig-bin}-(a), when DFT is performed on the dechirped signal $\boldsymbol{y}_{n,\ell}\otimes\boldsymbol{s}_0^*$, all the power of the chirp ${\boldsymbol{s}}_{\tilde{m}}$ is focused on the $\tilde{m}$th bin:
\begin{align}\label{fft}
	V_{n,\ell}[k]
	=\begin{cases}
		\sqrt{M}{h}_{u,\ell}\exp\left(j\psi_{\tilde{m}}\right) + W_{n,\ell}[k],\; k={\tilde{m}}\\
		W_{n,\ell}[k],\quad \text{otherwise},
	\end{cases}
\end{align}
where $
{\boldsymbol{W}}_{n,\ell}=[{{W}}_{n,\ell}[0],{{W}}_{n,\ell}[1]\dots {{W}}_{n,\ell}[M-1]]^{\T} =\mathrm{DFT}\left\{\boldsymbol{\omega}_{n,\ell}\right\},
$
and $
\theta_{\tilde{m}}=2\pi\left(\frac{ {\tilde{m}}^2}{2M}-\frac{{\tilde{m}}}{2}\right)$ is a deterministic bin-dependent phase offset, which can be conveniently discarded by converting $V_{n,\ell}[k]$ into:
%%%%%%%%%%%%%%%%%%
\begin{align}\label{fft2}
	\tilde{V}_{n,\ell}[k]= \frac{{V}_{n,\ell}[k]}{\exp\left(j\theta_k\right)}
	=\begin{cases}
		\sqrt{M}{h}_{u,\ell} + \tilde{W}_{n,\ell}[k],\; k={\tilde{m}}\\
		\tilde{W}_{n,\ell}[k],\quad \text{otherwise}.
	\end{cases}
\end{align}
Thus, when only one ED is transmitting, once the preamble detection and synchronization are finished, the demodulation can be done with square-law combining \cite{Nguyen2021}, by simply calculating the power of all $M$ frequency bins gathered from all $L$ antennas $\boldsymbol{\Upsilon}_n=\left[\Upsilon_{n}[0],\Upsilon_{n}[1]\dots \Upsilon_{n}[M-1]\right]^{\T}$, where:
\begin{equation}\label{eq.binP}
	\Upsilon_{n}[k] = \sum_{\ell=1}^{L}\left\vert \tilde{V}_{n,\ell}[k]\right\vert^2.
\end{equation}
The bin with the highest power is the one corresponding to the transmitted LoRa symbol.

Whereas, if $\boldsymbol{y}_{n,\ell}$ is not synchronized with a LoRa symbol, but overlapped between two consecutive LoRa symbols $\boldsymbol{s}_{m_1}$ and $\boldsymbol{s}_{m_2}$, such that $	\boldsymbol{y}_{n,\ell} = {h}_{i,\ell}\bar{\boldsymbol{s}}$, where 
\begin{equation}
	\bar{\boldsymbol{s}}= [\underbrace{s_{m_1,\tau+1},s_{m_1,\tau+2}\dots s_{m_1,M}}_{\text{tail of $\boldsymbol{s}_{m_1}$}},\underbrace{ s_{m_2,1}, s_{m_2,1}\dots s_{m_2,\tau}}_{\text{head of $\boldsymbol{s}_{m_2}$}}],
\end{equation}
with $\tau<M$.
When $\bar{\boldsymbol{s}}$ is dechirped by multiplying with $\boldsymbol{s}_0^*$, following \eqref{eq7}, the  signal $\bar{\boldsymbol{s}}\otimes\boldsymbol{s}_0^*$ has the $m$th elements as:

\begin{align}
	%\begin{split}
	&\bar{{s}}_m{s}_{0,m} =\begin{cases}
		s_{m_1,\tau+m}{s}_{0,m}^*,\;\;\quad\quad  m\leq M-\tau\\
		s_{m_2,m-(M-\tau)}{s}_{0,m}^*,\;\; m>M-\tau
	\end{cases}\nn\\
	&=\begin{cases}
		\exp\{j\theta_{\tilde{m}_1}\}\exp\left\{\frac{j2\pi \tilde{m}_1(m-1)}{M}\right\}, \; m\leq M-\tau\\
		\exp\{j\theta_{\tilde{m}_2}\}\exp\left\{\frac{j2\pi \tilde{m_2}(m-1)}{M}\right\}, \;m>M-\tau.
	\end{cases}
	%\end{split}
\end{align}
where $\tilde{m}_1=m_1-\tau$ and $\tilde{m}_2=m_2+M-\tau$. The dechirped signal now contains  fragments of two discrete sinusoids with two different frequencies. As a result,
the power in the two fragments from two consecutive LoRa symbols is spread unevenly over $M$ frequency bins, focused around the two frequencies $\tilde{m_1}$ and $\tilde{m}_2$ in the { frequency} domain as illustrated in Fig.~\ref{fig-bin}-(b), (c) and (d) with different values of $\tau$. 	
%			%%%%%%%%%%%
%\begin{IEEEeqnarray}{rcl}\label{eq6}
%	&&V_{n,\ell}[k] \nonumber\\
%	&&=\frac{1}{\sqrt{M}}\sum_{n=0}^{M-\tau}{\exp\left(j\Psi_m\right)}{\exp\left(\frac{j2\pi mn}{M}\right)} \nonumber\\
%	&&+\frac{1}{\sqrt{M}}\sum_{n=0}^{\tau}{\exp\left(j\Psi_m\right)}{\exp\left(\frac{j2\pi mn}{M}\right)} \nonumber\\
%\end{IEEEeqnarray}	
This highlights the importance of detecting the LoRa preamble and synchronization. 

\begin{figure}[t!]
	\centering
	\includegraphics[width=0.5\textwidth]{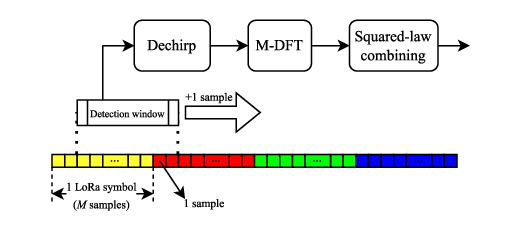}
	\caption{Conventional single-chirp preamble detection.}
	%\label{fig.2}
	\label{fig-preamble-detection}
\end{figure}

\begin{figure}[t!]
	\centering
	\includegraphics[width=0.5\textwidth]{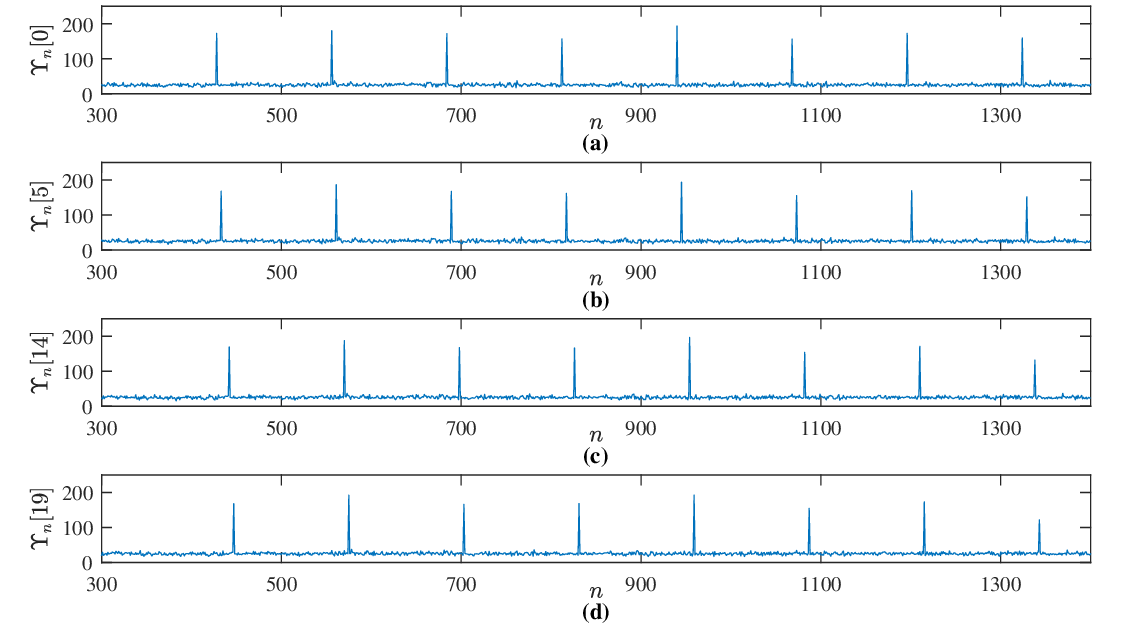}
	\caption{DFT output of the dechirped  window after squared-law combining over  time ($\mathrm{SF}=7$, $L=64$ antennas). The preamble of an ED can trigger PPs on every { frequency} bin (the 0th, 5th, 14th and 19th bins are arbitrarily chosen for  example. Any $m$th bin experiences the same phenomenon).}
	%\label{fig.1}
	\label{fig-peaks-2}
\end{figure}

Next, we review the preamble detection of the conventional LoRa to detect the preamble of a single ED at a time.
Since a LoRa symbol is $M$ samples in length,  to detect the preamble, the GW uses a window spanning over $M$ samples in the time-domain as shown in Fig.~\ref{fig-preamble-detection}. In \cite{Ghanaatian2019,Tang2019,Edward2019,bernier2020}, the window  slides along the time domain, one \emph{symbol} at a time, which has a low complexity. Another equivalent way is to shift the window by one \emph{sample} at a time. Although in this way, the complexity is higher since the signal processing must be performed every sample, rather than every symbol ($M$ samples), we chose to discuss the conventional preamble detection in this way since it is similar to the signal processing in our proposed design. The higher complexity issue (as compared to shifting the window every symbol) will be addressed in Section~\ref{Sec:com}. Conventionally, the LoRa preamble is made of $N=8$ consecutive basic upchirps ($\boldsymbol{s}_0$). To detect these $N$ consecutive basic upchirps, the GW observes the power of the  frequency bins in the frequency domain as in \eqref{eq.binP}.
Since the preamble is $N$ consecutive basic upchirps, the GW specifically observes the $0$th frequency bin $\Upsilon_{n}[0]$ over time to find  $N=8$ consecutive power peaks, separated by $M$ samples (or 1 LoRa symbol time) as shown in Fig.~\ref{fig-peaks-2}-a. To differentiate this with peak samples caused by noise or interference, we specifically refer to the consecutive peaks associated with the $N$ identical symbols of the preamble as \emph{preamble peaks} (PPs). Thus, when the GW observes $N$ consecutive basic upchirps, or in other words, $N$ consecutive PPs on $\Upsilon_{n}[0]$, it decides that a preamble is present, and a new packet is coming. Then, the GW synchronizes the reception to the time instance of the first detected peak with the aid of the SFD field \cite{Vangelista2022}.

%		\begin{figure}[t!]
	%			\centering
	%			\includegraphics[width=0.5\textwidth]{bin-preamble}
	%			\caption{FFT output of the dechirped FFT window over discrete time ($SF=7$, $L=64$ antennas, $\mathrm{SNR}=-20$ dB).}
	%			%\label{fig.1}
	%			\label{fig-peaks}
	%		\end{figure}

%\vspace{-0.4cm}
\subsection{Challenges with conventional { single-chirp} preamble  in multiuser LoRa}

%		In \cite{Kang2024}, the author propose a channel estimation algorithm based on preambles for multiuser LoRa networks. In specific, each user is assigned with a known preamble, and all the preamble are are mutually orthogonal. These preambles act as pilot sequences for channel estimation. This approach appears very straightforward and effective. For both preamble detection and channel estimation of an ED, the GW only needs to observe the frequency bin corresponding to that ED's designated preamble. However, it requires that all preambles are transmitted simultaneously and perfectly synchronized. The requirement means all EDs must follow a strict synchronization transmission at packet level. Given the nature of ALOHA used in LoRaWAN, and the fact the size of a LoRa packet is variable, this condition appears impractical.
In a multiuser scenario,  with the conventional preamble design with $N$ consecutive basic upchirps, once the GW detects the presence of a preamble, it is impossible to determine to which ED  the detected preamble belongs, based solely on the information of the preamble, since all EDs use the same preamble. Furthermore, under the influence of noise and interference among EDs, the probability that the GW fails to detect  preambles is high.

One possible solution is to assign different single-chirps to different EDs to use as preambles, for example, the $u$th ED is assigned with the $u$th chirp ($\boldsymbol{s}_u$) to use as a preamble. With this approach, although one can say that the GW only needs to observe the $u$th bin corresponding to the preamble of the $u$th ED to detect its incoming packet, this task is far from simple. This is because, as the detection window slides in the time domain to capture $M$ consecutive samples, the preamble of the $u$th ED (assumed to be a repetition of the $u$th chirp)  results not only in $N$ PPs in the $u$th bin, but it also creates $N$ PPs in every other  frequency bin. This effect is illustrated in Fig.~\ref{fig-peaks-2},
and  can be explained as follows. Suppose the first preamble symbol of the $u$th ED  ($\boldsymbol{s}_u$) is synchronized completely into the detection window at time $n$, which is:
%\vspace{-.24cm}
\begin{equation}
	\boldsymbol{y}_{n,\ell} = {h}_{u,\ell}{\boldsymbol{s}}_{u}+\boldsymbol{\omega}_{n,\ell}, 
\end{equation}
and results in  a PP in the corresponding $u$th bin. Since the chirp $\boldsymbol{s}_u$ is repeated $N$ times, when the detection window shifts by one sample, the signal $ \boldsymbol{y}_{n+1,\ell}$  taken from the discrete time $n+1$ to $n+M$ will capture the last $M-1$ samples of the chirp { $\boldsymbol{s}_u$} that belongs to the first symbol of the preamble, and the first sample of the second symbol of the preamble  $s_{u,1}$, which is exactly the $(u+1)$th chirp $\boldsymbol{s}_{u+1}$.  As a result:
\begin{equation}
	\begin{split}
		\boldsymbol{y}_{n+1,\ell} &= {h}_{u,\ell}[s_{u,2}, s_{u,2},\dots s_{u,M}, s_{u,1}]^{\T}+\boldsymbol{\omega}_{n+1,\ell}\\
		& = {h}_{u,\ell}\boldsymbol{s}_{u+1}+\boldsymbol{\omega}_{n+1,\ell} ,
	\end{split}
\end{equation}
which  results in a PP on the $(u+1)$th bin. Similarly, it can be seen that at time $n+\tau$, there will be a PP  on the $(u+\tau)$th bin, which explains the effect illustrated in Fig.~\ref{fig-peaks-2}.
Consequently, once a preamble is transmitted, every bin in the DFT domain will  experience $N$ consecutive peaks as shown in Fig. \ref{fig-peaks-2}, where a preamble of $N=8$ consecutive basic upchirps is transmitted. From the figure, it can be seen that the preamble triggers $N$ PPs on not only the $u$th bin, but also all other bins following it. In this case, with multiple EDs, the GW cannot determine which ED is transmitting the packet, even though every ED has its own unique preamble. 

The above discussion shows that the biggest problem in preamble detection of multiuser LoRa networks is what we call \emph{preamble resemblance}, which means that the preamble of an ED can undesirably resemble the preamble of others. Thus, it is important to develop a novel preamble design  to address this problem in  multiuser  LoRa networks.

%This effect, when considering only one preamble transmitted, has already been able to confuse the GW in preamble detection. When resonating with multiple preambles transmitted simultaneously, it will fail the operation of LoRa networks completely.
\vspace{-.1cm}
\section{Novel double-chirp preamble design for multiuser  MIMO LoRa}\label{Sec:preamble}

In this section, a novel preamble design for multiuser massive MIMO LoRa networks is proposed. This preamble design  helps to avoid preamble resemblance, which enables multiuser preamble detection in LoRa networks.
\vspace{-.5cm}
\subsection{Double-chirp preamble design for multiuser  MIMO LoRa }
As shown in Fig.~\ref{fig-peaks-2}, even when all EDs are assigned with unique \emph{single-chirp} preambles,  the preamble of any ED can trigger $N$ PPs in every { frequency bin}, which resembles the preambles of all others, and causes failure to the preamble detection of multiuser LoRa networks. Thus, instead of using \emph{single-chirp} preamble, we propose to assign to each ED  a {\em unique double-chirp} preamble. Specifically, the preamble of an arbitrary ED is the sum of two different chirps that are uniquely assigned to this ED, repeated $N$ times. Suppose that the $u$th ED is assigned  two chirps $\boldsymbol{s}_{\kappa_{u,1}}$ and $\boldsymbol{s}_{\kappa_{u,2}}$ to use as a preamble, then, the transmitted preamble of the $u$th ED is:
%%%%%%%%%%%%%%%%%
\begin{equation}
	\boldsymbol{\rho}_u =\frac{1}{\sqrt{2}} (\boldsymbol{s}_{\kappa_{u,1}}+\boldsymbol{s}_{\kappa_{u,2}}),
\end{equation}
%%%%%%%%%%%%%%%%%%
repeated $N$ times.
To detect the preamble of this ED, the GW still uses the conventional DFT-based detector  to calculate the signal of all $M$  bins  on all $L$ antennas as in \eqref{eq-pdw-V}. Suppose the preamble  $\boldsymbol{\rho}_u$ from the $u$th ED is received at time $n$:
\begin{equation}
	\boldsymbol{y}_{n,\ell} = {h}_{u,\ell}{\boldsymbol{\rho}}_{u}+\boldsymbol{\omega}_{n,\ell}.
\end{equation}
Since $\boldsymbol{\rho}_u$ is the sum of the two upchirps $\boldsymbol{s}_{\kappa_{u,1}}$ and $\boldsymbol{s}_{\kappa_{u,2}}$, the signals on $M$ frequency bins are:
%%%%%%%%%%%%%%%%%%
\begin{align}\label{fft1}
	\tilde{V}_{n,\ell}[k]
	=\begin{cases}
		\sqrt{\frac{M}{2}}{h}_{u,\ell}+ 	\tilde{W}_{n,\ell}[k],\; k=\kappa_{u,1} \text{ or } \kappa_{u,2}\\
		0,\quad \text{otherwise}.
	\end{cases}
\end{align}
%%%%%%%%%%%%%%%%
Let $\check{\boldsymbol{V}}_{n,k}=\left[\tilde{V}_{n,1}[k], \tilde{V}_{n,2}[k],\dots \tilde{V}_{n,L}[k]\right]^{\T}\in\mathbb{C}^{L\times1}$ be the signal vector collected on all $L$ antennas in the $k$th bin at time $n$, the above equation can be rewritten in vector form as:
%%%%%%%%%%%%%
\begin{equation}\label{fft1v}
	\check{\boldsymbol{V}}_{n,k}
	=\begin{cases}
		\sqrt{\frac{M}{2}}\boldsymbol{h}_{u}+ 	\check{\boldsymbol{W}}_{n,k},\quad k=\kappa_{u,1} \text{ or } \kappa_{u,2}\\
		\check{\boldsymbol{W}}_{n,k},\quad \text{otherwise},
	\end{cases}
\end{equation}
%%%%%%%%%%%%%%%%
where $\check{\boldsymbol{W}}_{n,k}=\left[\tilde{W}_{n,1}[k], \tilde{W}_{n,2}[k],\dots \tilde{W}_{n,L}[k]\right]^{\T}\in\mathbb{C}^{L\times1}$ is the noise vector collected on all $L$ antennas in the $k$th bin at time $n$.
%$	\boldsymbol{y}_{n,\ell} = \sqrt{\rho_i}{h}_{i,\ell}{\boldsymbol{s}}_{m}$
Thus, rather than performing a square-law combining  and tracking the power of the corresponding bins of that ED as for the single ED case, the GW combines the signals over $L$ antennas of the two  bins $\check{\boldsymbol{V}}_{n,\kappa_{u,1}}$ and $\check{\boldsymbol{V}}_{n,\kappa_{u,2}}$ of the $u$th ED:
%%%%%%%%%%%%%%%
%%%%%%%%%%%%%%%%%%
%\begin{figure*}[h!]
%\begin{eqnarray}
\begin{align}
	%\begin{split}
	\label{eq-pdw-Z-PP}
	&{Z}_{n,u}
	= \frac{1}{L}\mathfrak{R}\left\{ \check{\boldsymbol{V}}_{n,\kappa_{u,1}}^{\Hi}\check{\boldsymbol{V}}_{n,\kappa_{u,2}}\right\}\nn\\
	& = \frac{1}{L}\mathfrak{R}\left\{\left(\sqrt{\frac{M}{2}}\boldsymbol{h}_u + \check{\boldsymbol{W}}_{n,\kappa_{u,1}}\right)^{\Hi}\left(\sqrt{\frac{M}{2}}\boldsymbol{h}_u + \check{\boldsymbol{W}}_{n,\kappa_{u,2}}\right)\right\}\nn\\
	& =  \frac{1}{L}\mathfrak{R}\left\{\sqrt{\frac{M}{2}}\boldsymbol{h}_u^{\Hi}\left(\check{\boldsymbol{W}}_{n,\kappa_{u,1}}+\check{\boldsymbol{W}}_{n,\kappa_{u,2}}\right)+
	\check{\boldsymbol{W}}_{n,\kappa_{u,1}}^{\Hi}\check{\boldsymbol{W}}_{n,\kappa_{u,2}}\right\}\nn\\
	&+\frac{M}{2L}\Vert\boldsymbol{h}_u\Vert^2.
	%\xrightarrow[M\rightarrow\infty]{\mathrm{a.s}}\frac{M}{2},\nn
	%\end{split}
\end{align}
For the remainder of this paper, ${Z}_{n,u}\in\mathbb{R}$ is referred to as the bin-combined preamble (BCP) signal of the $u$th ED.
According to the law of large numbers \cite{Marzetta2016}, the term
%%%%%%%%%%%%%%%
\begin{align}\label{eq20}
	%\begin{split}
	\frac{1}{L}\mathfrak{R}&\left\{\sqrt{\frac{M}{2}}\boldsymbol{h}_u^{\Hi}\left(\check{\boldsymbol{W}}_{n,\kappa_{u,1}}+\check{\boldsymbol{W}}_{n,\kappa_{u,2}}\right)+
	\check{\boldsymbol{W}}_{n,\kappa_{u,1}}^{\Hi}\check{\boldsymbol{W}}_{n,\kappa_{u,2}}\right\}\nn\\
	&\xrightarrow[L\rightarrow\infty]{\mathrm{a.s}}0,
	%\end{split}
\end{align}
%%%%%%%%%%%%%%%
where $\xrightarrow[L\rightarrow\infty]{\mathrm{a.s}}$ denotes almost sure convergence when the number of antennas $L$ tends to infinity \cite{Marzetta2016}.
This is because \eqref{eq20} consists of terms which are the multiplication between two statistically independent vectors \cite{Marzetta2016}. On the other hand, also resulted from the law of large numbers
\begin{equation}\label{eq.chan}
	\frac{M}{2L}\Vert\boldsymbol{h}_u\Vert^2 
	\xrightarrow[L\rightarrow\infty]{\mathrm{a.s}}\frac{M}{2}.
\end{equation}
As a result
\begin{equation}
	Z_{n,u}
	\xrightarrow[L\rightarrow\infty]{\mathrm{a.s}}\frac{M}{2}.
\end{equation}
Whereas, any other combination of two bins other than between the $\kappa_{u,1}$th and $\kappa_{u,2}$th bins is just the multiplication between two independently distributed noise vectors, which will result in zero convergence
\begin{equation}
	\begin{split}
		\label{eq-pdw-Z}
		& \frac{1}{L}\mathfrak{R}\left\{ \check{\boldsymbol{V}}_{n,k_1}^{\Hi}\check{\boldsymbol{V}}_{n,k_2}\right\}=  \frac{1}{L}\mathfrak{R}\left\{
		\check{\boldsymbol{W}}_{n,k_1}^{\Hi}\check{\boldsymbol{W}}_{n,k_2}\right\}\xrightarrow[L\rightarrow\infty]{\mathrm{a.s}}0,
		%\xrightarrow[M\rightarrow\infty]{\mathrm{a.s}}\frac{M}{2}\nn
	\end{split}
\end{equation}
if $\left\{k_1,k_2\right\}\neq\left\{\kappa_{u,1},\kappa_{u,2}\right\}$ \cite{Marzetta2016}. Consequently, the \emph{double-chirp} preamble $\boldsymbol{\rho}_u$ will not trigger any resembled PPs on the BCP signal $Z_{n,\tilde{u}}$ ($\tilde{u}\neq u$) of any other EDs at time $n$.
Thus, this preamble design can reduce the chance that the preamble of the $u$th ED triggers the bins corresponding to the preambles of other EDs and causes preamble resemblance.

However, this preamble design still cannot completely get rid of  the preamble resemblance effect that can make the GW make wrong decision on the preamble detection. In the following, Section \ref{Sec:preamble}-B will study the preamble resemblances that are caused by  {\em single-ED} sources, and how they can be eliminated with { careful} preamble assignment. Next, Section \ref{Sec:preamble}-C will discuss the preamble resemblance that is jointly caused by {\em two different EDs}, which cannot be removed with preamble assignment, but can still be recognized by the GW.
\subsection{Preamble resemblance from single-ED sources and the solution by preamble assignment}
This subsection discusses the preamble resemblance of the proposed {\em double-chirp} preamble design that is caused by a \emph{single ED}, and how to address it with preamble assignment. 
\subsubsection{Inter-ED preamble resemblance}
%Similar to the case of single-ED LoRa networks,  the preamble in multiuser scenario is also repeated $N$ times. 
Suppose the $u$th ED is assigned with chirps $\boldsymbol{s}_{\kappa_{u,1}}$ and $\boldsymbol{s}_{\kappa_{u,2}}$ as preamble, and the $\tilde{u}$th ED is assigned with $\boldsymbol{s}_{1+\kappa_{u,1}}$ and $\boldsymbol{s}_{1+\kappa_{u,2}}$. At time $n$, if
\begin{equation}
	\boldsymbol{y}_{n,\ell} = {h}_{u,\ell}{\boldsymbol{\rho}}_{u}+\boldsymbol{\omega}_{n,\ell}={h}_{m,\ell}\frac{1}{\sqrt{2}}(\boldsymbol{s}_{\kappa_{u,1}}+\boldsymbol{s}_{\kappa_{u,2}})+\boldsymbol{\omega}_{n,\ell}, 
\end{equation}
then, similar to the single ED case in the previous section, at time $n+1$, when the detection window shifts by one sample
\begin{equation}
	\begin{split}
		\boldsymbol{y}_{n+1,\ell} &= {h}_{u,\ell}\frac{1}{\sqrt{2}} (\boldsymbol{s}_{1+\kappa_{u,1}}+\boldsymbol{s}_{1+\kappa_{u,2}})+\boldsymbol{\omega}_{n+1,\ell}\\
		&={h}_{u,\ell}{\boldsymbol{\rho}}_{\tilde{u}}+\boldsymbol{\omega}_{n+1,\ell}, 
	\end{split}
\end{equation}
which will trigger an unwanted PP in { ${Z}_{n,\tilde{u}}$} for the $\tilde{u}$th ED at time $n+1$. Similarly, if another ED is assigned with  $\boldsymbol{s}_{\tau+\kappa_{u,1}}$ and $\boldsymbol{s}_{\tau+\kappa_{u,2}}$ as preamble, it will also get an unwanted PP at time $n+\tau$ if the preamble of the first ED is received at time $n$. This effect will be illustrated in the following example.

%\end{eqnarray}
%\end{figure*}
%%%%%%%%%%%%%%%%%%%
%%%%%%%%%%%%%%%
\begin{figure}[t!]
	\centering
	\includegraphics[width=0.5\textwidth]{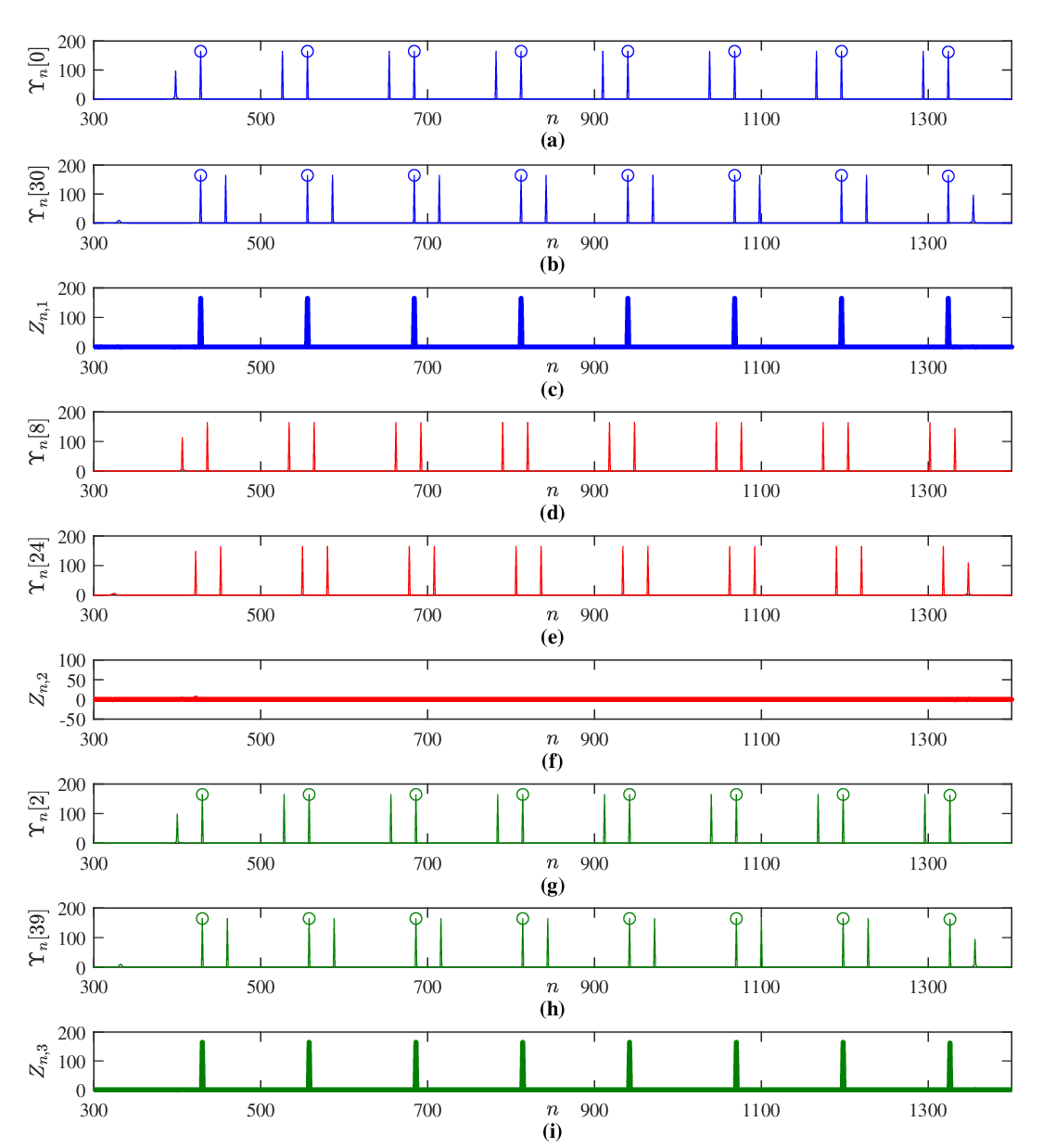}
	\caption{Preamble signal of the first ED (blue) does not resemble the second ED (red), but resembles the third ED (green) ($\mathrm{SF}=7$, $L=64$, $\Delta_1=\Delta_3 = 30$, $\Delta_2=16$). Circles indicate  two different bins triggered at the same time, leading to PPs in the BCP signal of an ED.}
	%\label{fig.1}
	\label{fig-peaks-single_resemblance}
\end{figure}

	\textbf{Example 1:} This example illustrates how the proposed novel preamble design can enable preamble detection of an ED, and when preamble resemblance can occur.
	
	Consider a LoRa network with 3 EDs and a GW with $L=64$ antennas. Suppose the first ED is assigned with 2 chirps $\boldsymbol{s}_0$ and $\boldsymbol{s}_{30}$ as preamble,  the second ED is assigned with $\boldsymbol{s}_8$ and $\boldsymbol{s}_{24}$, and the third ED is assigned with $\boldsymbol{s}_2$ and $\boldsymbol{s}_{32}$. Assume that only the first ED is transmitting its preamble with length $N=8$. The power of the DFT output on the bins corresponding to the aforementioned chirps (before combining) over \emph{discrete time} is illustrated in Fig~\ref{fig-peaks-single_resemblance}-(a), (b), (d), (e), (g), (h). Since the first ED uses both $\boldsymbol{s}_0$ and $\boldsymbol{s}_{30}$ as preamble,  $2N=16$ PPs in total appear on every bin $\Upsilon_{n}[k]$ ($k=0,1\dots M-1$) when performing square-law combining.
	
	However, when combining the $0$th and the $30$th bins to obtain $Z_{n,1=}=\check{\boldsymbol{V}}_{n,0}^{\Hi}\check{\boldsymbol{V}}_{n,30}$, which the the BCP signal  of the first ED (in  Fig~\ref{fig-peaks-single_resemblance}-(c)), only $N=8$ PPs remain. The reason is that $\boldsymbol{s}_0$ and $\boldsymbol{s}_{30}$ are transmitted at the same time by the first ED. As a result, these two chirps trigger their own PPs on the corresponding bins ($0$th and $30$th bins) at the same instant, and hence, the combined signal from these two bins obtained with \eqref{eq-pdw-Z} also has PPs at the same position (marked by blue circles in  Fig~\ref{fig-peaks-single_resemblance}-(a), (b)). On the other hand, the 8 PPs that $\boldsymbol{s}_0$ triggers on the $30$th bin, and the 8 PPs that $\boldsymbol{s}_{30}$ triggers on the $0$th bin are not aligned. Thus, these unwanted peaks will not appear on the BCP signal ${Z}_{n,1}$ of the first ED.
	
	Similarly, the PPs that the preamble of the first ED triggers on the $8$th and $24$th bins are not aligned either (Fig~\ref{fig-peaks-single_resemblance}-(d), (e)). Thus, the BCP signal of the second ED ${Z}_{n,2}=\check{\boldsymbol{V}}_{n,8}^{\Hi}\check{\boldsymbol{V}}_{n,24}$ does not have any PPs resembled by the preamble of the first ED, as shown in Fig~\ref{fig-peaks-single_resemblance}-(f).
	Therefore, with this design, the preamble of the first ED can be detected by the GW without resembling the preamble  of the second ED. 
	
	However,  the BCP signal ${Z}_{n,3}=\check{\boldsymbol{V}}_{n,2}^{\Hi}\check{\boldsymbol{V}}_{n,32}$  of the third ED experiences $N=8$ consecutive peaks similar to that of the first ED as shown in Fig.~\ref{fig-peaks-single_resemblance}-(i).  The reason is that after the PPs of the first ED is detected, when the detection window shifts in the time domain by two samples, $\boldsymbol{s}_0$ becomes $\boldsymbol{s}_2$ and $\boldsymbol{s}_{30}$ becomes $\boldsymbol{s}_{32}$, which trigger the peaks in the second and the 32th bins at the same time as marked in circle in Fig.~\ref{fig-peaks-single_resemblance}-(g), (h), resulting in the PPs in the BCP signal ${Z}_{n,3}$ of the third ED as in Fig.~\ref{fig-peaks-single_resemblance}-(i). As a consequence, the preamble of the first ED appears identical to the preamble of the third ED, two samples after it was detected for the first ED. $\hfill\blacksquare$

Fortunately, this problem can be easily solved. Since the preamble constructed from the two chirps $\boldsymbol{s}_{\kappa_{u,1}}$ and $\boldsymbol{s}_{\kappa_{u,2}}$ will also resemble the preamble of any ED that uses $\boldsymbol{s}_{\kappa_{u,1}+\tau}$ and $\boldsymbol{s}_{\kappa_{u,2}+\tau}$ as preamble, the undesirably PPs can be avoided by assigning preamble chirps to EDs such that the distance:
\begin{equation}\label{eq.Delta}
	\Delta_u=\mathrm{mod}\left(\left\vert\kappa_{u,1}-\kappa_{u,2}\right\vert,M/2\right),
\end{equation}
is unique for each and every ED in the network. Since the LoRa chirp set is periodic with the period of $M$ samples, this condition is generalized as
\begin{equation}\label{eq.con2}
	\Delta_u\neq\Delta_{\tilde{u}},\;\forall u,\tilde{u}.
\end{equation} 

\subsubsection{Self preamble resemblance}
\begin{figure}[t!]
	\centering
	\includegraphics[width=0.5\textwidth]{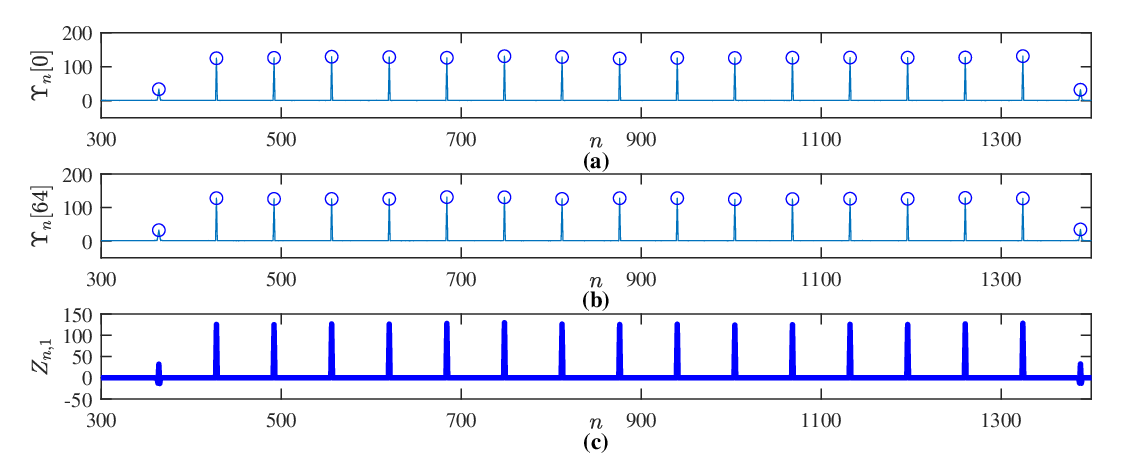}
	\caption{Self preamble resemblance when an ED is assigned with $\boldsymbol{s}_0$ and $\boldsymbol{s}_{64}$ as preamble. $2N=16$ PPs appear instead of $N=8$ PPs ($\Delta_m=0$, $\mathrm{SF}=7$, $L=64$). Circles indicate  two different bins triggered at the same time, leading to PPs in the BCP signal.}
	%\label{fig.1}
	\label{fig-peaks-3}
\end{figure}

Unlike inter-ED preamble resemblance, where an ED's preamble can resemble that of other EDs, self preamble resemblance is caused by an ED to itself. This happens when the $u$th ED is assigned with two chirps $\boldsymbol{s}_{\kappa_{u}}$ and $\boldsymbol{s}_{\kappa_{u}+M/2}$  as preamble. Suppose at time $n$, if
\begin{equation}
	\boldsymbol{y}_{n,\ell} = {h}_{u,\ell}{\boldsymbol{\rho}}_{u}+\boldsymbol{\omega}_{n,\ell}={h}_{u,\ell}\frac{1}{\sqrt{2}}(\boldsymbol{s}_{\kappa_{u}}+\boldsymbol{s}_{\kappa_{u}+M/2})+\boldsymbol{\omega}_{n,\ell}, 
\end{equation}
then, similar to the explanation in previous subsection on the inter-ED case, at time $n+M/2$
\begin{equation}
	\boldsymbol{y}_{n+M/2,\ell} = {h}_{u,\ell}\frac{1}{\sqrt{2}} (\boldsymbol{s}_{\kappa_{u}+M/2}+\boldsymbol{s}_{\kappa_{u}})+\boldsymbol{\omega}_{n+M/2,\ell}, 
\end{equation}
which  triggers a resembled PP of the same $u$th ED. Consequently, instead of having $N$ consecutive PPs, separated by $M$ samples, the BCP signal $Z_{n,u}$ of the $u$th ED will experience $2N$ PPs, separated by $M/2$ samples as shown in Fig.~\ref{fig-peaks-3}.

As can be seen, this phenomenon only occurs when the double-chirp preamble has $\left\vert\kappa_{u,1}-\kappa_{u,2}\right\vert=M/2$. Thus, to avoid this problem, in addition to the condition in {  \eqref{eq.con2}}, the double-chirp preamble must also satisfy:
\begin{equation}\label{eq.con1}
	\Delta_u\neq 0,\;\forall u.
\end{equation}
%\vspace{-.5cm}
\subsubsection{ED limit with preamble assignment} By the definition  in \eqref{eq.Delta}, and with the constraint in \eqref{eq.con1}, $\Delta_u$ must satisfy $0<\Delta_u<M/2$. Since all $\Delta_u$ must be unique according to \eqref{eq.con2}, the maximum number of EDs that can be supported without \emph{single-ED} preamble resemblance is technically $(M/2-1)$. 
\subsection{Preamble resemblance from joint-ED source}
In the previous subsection, the preamble resemblance caused by a single ED has been investigated. Fortunately, with the proposed preamble assignment method, the effect that the preamble of a single ED resembles the  preambles  of other EDs (and/or itself) can be avoided. However, this only prevents an ED from resembling, \emph{single-handedly},  the preamble of other EDs. In this subsection, we examine another case of preamble resemblance  that involves two different EDs \emph{jointly} resembling the preamble of a third ED.

Without  loss of generality, suppose the first ED transmits the preamble $\boldsymbol{\rho}_1= [(\boldsymbol{s}_{\kappa_{1,1}}+\boldsymbol{s}_{\kappa_{1,2}})]/\sqrt{2}$ at time $n$ and the second ED transmits the preamble	$\boldsymbol{\rho}_2= [(\boldsymbol{s}_{\kappa_{2,1}}+\boldsymbol{s}_{\kappa_{2,2}})]/\sqrt{2}$ at time $n+\tau$ (where $\tau<M$), we have:
\begin{equation}
	\begin{split}
		\boldsymbol{y}_{n+\tau,\ell} 
		&=\frac{{h}_{1,\ell}(\boldsymbol{s}_{\kappa_{1,1}+\tau}+\boldsymbol{s}_{\kappa_{1,2}+\tau})+{h}_{2,\ell}(\boldsymbol{s}_{\kappa_{2,1}}+\boldsymbol{s}_{\kappa_{2,2}})}{\sqrt{2}}+\boldsymbol{\omega}_{n,\ell}, \\
	\end{split}
\end{equation}
As a result, at time $n+\tau$:
\begin{align}\label{fft3}
	\tilde{V}_{n+\tau,\ell}[k]
	=\begin{cases}
		\sqrt{\frac{M}{2}}{h}_{1,\ell}+\tilde{W}_{n+\tau,\ell}[k],\; k\in\left\{\kappa_{1,1}+\tau, \kappa_{1,2}+\tau\right\}\\
		\sqrt{\frac{M}{2}}{h}_{2,\ell}+\tilde{W}_{n+\tau,\ell}[k],\; k\in\left\{\kappa_{1,1}, \kappa_{1,2}\right\}\\
		\tilde{W}_{n+\tau,\ell}[k],\quad \text{otherwise}.
	\end{cases}
\end{align}
Consequently, the preamble of these two EDs can resemble the preamble of a third ED that is assigned with $\boldsymbol{s}_{\kappa_{3,1}}$ and $\boldsymbol{s}_{\kappa_{3,2}}$ as preamble, if $\kappa_{3,1}$ and $\kappa_{3,2}$ take  the values in $\left\{\kappa_{1,1}+\tau, \kappa_{1,2}+\tau,\kappa_{2,1},\kappa_{2,2}\right\}$. In this case:
%			\begin{equation}
	%		\begin{split}
		%			\label{eq-pdw-Z2}
		%			{Z}_{n,3}
		%			&= \frac{1}{L}\mathfrak{R}\left\{\sum_{\ell=1}^{L}\left\vert \tilde{V}_{n,\ell}[\kappa_{3,1}] \times\left(\tilde{V}_{n,\ell}[\kappa_{3,2}]\right)^{*}\right\vert\right\}\\
		%			& = \frac{M}{2L}\mathfrak{R}\left\{\boldsymbol{h}_{1}^{\Hi}\boldsymbol{h}_2\right\}\xrightarrow[L\rightarrow\infty]{\mathrm{a.s}}0,\nn
		%		\end{split}
	%	\end{equation}
\begin{align}
	%\begin{split}
	\label{eq-pdw-Z-PP2}
	&{Z}_{n+\tau,3}
	= \frac{1}{L}\mathfrak{R}\left\{ \check{\boldsymbol{V}}_{n+\tau,\kappa_{3,1}}^{\Hi}\check{\boldsymbol{V}}_{n+\tau,\kappa_{3,2}}\right\}=\nn\\
	&  \frac{1}{L}\mathfrak{R}\left\{\left(\sqrt{\frac{M}{2}}\boldsymbol{h}_1 + \check{\boldsymbol{W}}_{n+\tau,\kappa_{3,1}}\right)^{\Hi}\left(\sqrt{\frac{M}{2}}\boldsymbol{h}_2 + \check{\boldsymbol{W}}_{n+\tau,\kappa_{3,2}}\right)\right\}\nn\\
	&\xrightarrow[L\rightarrow\infty]{\mathrm{a.s}}0,
	%\xrightarrow[M\rightarrow\infty]{\mathrm{a.s}}\frac{M}{2}\nn
	%\end{split}
\end{align}
since $\boldsymbol{h}_1$ and $\boldsymbol{h}_2$ are statistically independent, according to the law of large number \cite{Marzetta2016}.
%	\begin{equation}
	%		Z_{n+\tau,3}
	%		\xrightarrow[L\rightarrow\infty]{\mathrm{a.s}}0.
	%	\end{equation}	

Although $Z_{n+\tau,3}$ converges to zero when $L$ tends to  infinity, with a limited number of antennas, 	$Z_{n+\tau,3}$  still experiences preamble resemblance. This is due to the correlation between $\boldsymbol{h}_1$ and $\boldsymbol{h}_2$, which is independent of $n$, and results in the consecutive peaks, despite being of low amplitudes compared to  true PPs.  This effect will be shown in the next example.

\begin{figure}[t!]
	\centering
	\includegraphics[width=0.5\textwidth]{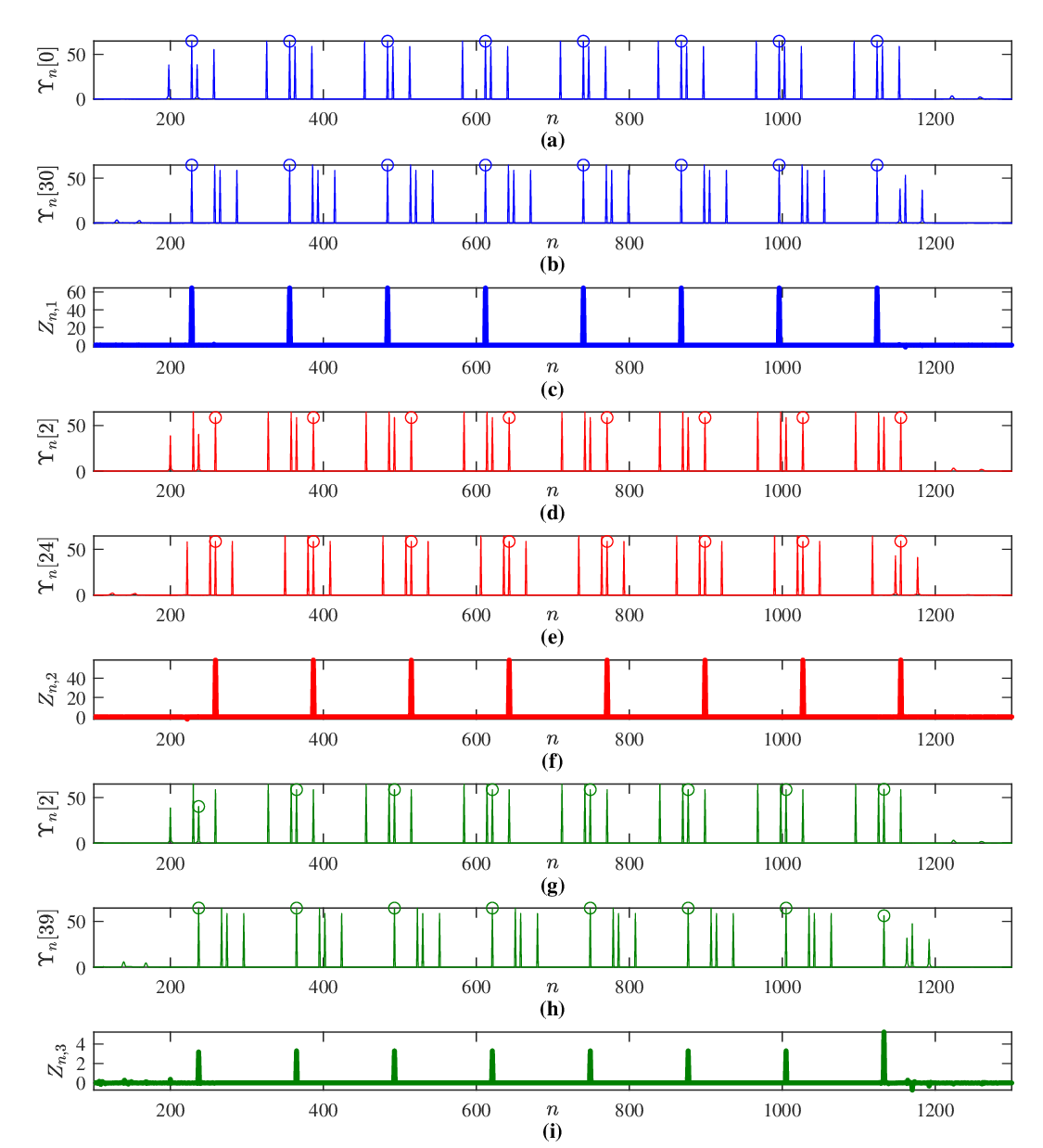}
	\caption{The first and the second EDs' preamble (blue and red, respectively) do not resemble the preamble of each another, but jointly resemble the preamble of the third ED (green) ($\mathrm{SF}=7$, $L=64$). Circles indicate  two different bins triggered at the same time, leading to PPs in the BCP signal of an ED.}
	%\label{fig.1}
	\label{fig-peaks-4}
\end{figure}

	\textbf{Example 2:} This example shows how two different EDs can resemble the preamble of a third ED. 
	
	Consider a LoRa network with 3 EDs and a GW with $L=64$ antennas. Suppose the first ED is assigned with 2 chirps $\boldsymbol{s}_0$ and $\boldsymbol{s}_{30}$ as preamble, and the second ED is assigned with $\boldsymbol{s}_2$ and $\boldsymbol{s}_{24}$, and the third ED is assigned with $\boldsymbol{s}_2$ and $\boldsymbol{s}_{39}$, with the respective DFT output over \emph{discrete time} (before combining) illustrated in Fig~\ref{fig-peaks-4}-(a), (b), (d), (e), (g), (h).
	Assume that the first and the second EDs are transmitting their preambles, while the third ED is idle. The time of arrival (ToA) of the first ED is $n= 0$ and the ToA of the second  is $n = 31$.
	
	The BCP signals of the first and the second EDs ${Z}_{n,1}$ and ${Z}_{n,2}$ are shown in Fig.~\ref{fig-peaks-4}-(c) and (f), respectively. As can be seen, with the double-chirp design with non-zero $\Delta_1\neq\Delta_2$, the preamble of each ED does not resemble the preamble of the other. Thus, the transmitted preambles of the first and  second EDs can be detected without preamble resemblance.
	
	However, while observing the BCP signal ${Z}_{n,3}$ of the third ED, the GW will also see consecutive PPs even though the third ED is idle as shown in Fig.~\ref{fig-peaks-4}-(i). The reason is that  at time $n=9$, the chirp $\boldsymbol{s}_{30}$ (belongs to the first ED) will become $\boldsymbol{s}_{39}$ (shifted 9 samples to the right). Whereas, the chirp $\boldsymbol{s}_{24}$ of the second ED will become $\boldsymbol{s}_{2}$ (shifted 9 samples to the right, then 31 samples to the left since the ToA of the second ED is 31 samples later than the first ED). Consequently, the PPs on the second and the 39th bins will be triggered at the same time (marked by green circles) as shown in Fig.~\ref{fig-peaks-4}-(g), (h). Thus, ${Z}_{n,3}=\check{\boldsymbol{V}}_{n,2}^{\Hi}\check{\boldsymbol{V}}_{n,32}$ will experience $N$ PPs as a result of the correlation between chirp $\boldsymbol{s}_{30}$ of the first ED and chirp $\boldsymbol{s}_{24}$ of the second ED. 
	
	From Fig.~\ref{fig-peaks-4}-(i), it can be seen that the amplitudes of these resembled PPs are of small values in comparison with the true PPs of the first and the second EDs, since the channels of the first and the second EDs are independent. However, they  can still potentially be mistaken as a true preamble.	$\hfill\blacksquare$

Unlike single-ED preamble resemblance that can be avoided with preamble assignment, joint-ED preamble resemblance cannot be. Its occurrence depends on the randomness of the difference between the ToAs of two EDs. Consequently, the resembled preamble caused by joint-ED accounts for  the majority of errors in preamble detection in multiuser massive MIMO LoRa networks. 
Fortunately, thanks to the statistical independence among channels of different EDs to the GW, the resembled PPs caused by joint-ED preamble resemblance are mostly of low amplitudes, and are equally likely to be either positive or negative. This results in the statistical difference between a resembled preamble and a true one, which has the PPs most certainly of positive amplitude with the mean of $M/2$ as shown in \eqref{eq-pdw-Z-PP}. Based on this statistical difference, in the next section, we will derive a maximum-likelihood (ML) based receiver, which determines the presence of an ED's preamble by measuring the likelihood of it being a true preamble, or a resembled one, or noise/interference.

\section{ML-based { Non-coherent} Detection for Double-chirp Preamble Design}\label{Sec:ML}
Since a preamble consists of $N$ consecutive identical symbols, in the presence of a new packet from the $u$th ED, the GW would expect to observe $N$ consecutive PPs, equally spaced by $M$ samples, when observing  ${Z}_{n,u,}$ as shown in Fig. \ref{fig-peaks-3}. Let:
%	\begin{equation}\label{eq.Z8}
	%		\begin{split}
		%			{\boldsymbol{Z}}_{n,u} =
		%			\begin{bmatrix}
			%				{Z}_{n-M(N-1),u}\\
			%				{Z}_{n-M(N-2),u}\\
			%				\vdots\\ 
			%				{Z}_{n-M,u}\\
			%				{Z}_{n,u} 		
			%			\end{bmatrix}\in\mathbb{C}^{N\times1},
		%		\end{split}
	%	\end{equation}
\begin{equation}\label{eq.Z8}
	{\boldsymbol{Z}}_{n,u} =
	[{Z}_{n-M(N-1),u},
	{Z}_{n-M(N-2),u},
	\dots 	
	{Z}_{n,u}]^{\T} \in\mathbb{C}^{N\times1}	,	
\end{equation}
which is obtained by taking $N$ samples, separated by $M$ samples, from the BCP signal $Z_{n,u}$ of the $u$th ED.	
Conventionally, the preamble can be detected with majority vote method \cite{Kang2024}, which means that when at least $N_{\mathrm{thr}}$ out of $N$ elements of the BCP signal $\boldsymbol{Z}_{n,u}$ have the magnitudes greater than a predetermined threshold, the GW will determine that the preamble of the $u$th is present.

%		\begin{equation}
	%			\begin{split}
		%				&f\left(\tilde{Z}_{t}[t]\text{ is PP}\right) = \frac{1}{\sqrt{2\pi\sigma_{\mathrm{PP}}^2}}\exp\left\{-\frac{\left(\tilde{Z}_{m}[t]-M/2\right)^2}{2\sigma_{\mathrm{PP}}^2}\right\}		,	
		%			\end{split}
	%		\end{equation}
%		and the likelihood of  $\tilde{Z}_{m}[t]$ being noise is:
%		\begin{equation}\label{eq.LLnoise}
	%			\begin{split}
		%				&f\left(\tilde{Z}_{m}[t]\text{ is noise}\right) = \frac{1}{\sqrt{2\pi\sigma_{\mathrm{n}}^2}}\exp\left\{-\frac{\left(\tilde{Z}_{m}[t]\right)^2}{2\sigma_{\mathrm{n}}^2}\right\}.
		%			\end{split}
	%		\end{equation}
%		Based on these two likelihood functions, each sample of $\check{\boldsymbol{Z}}_m[t] $ can be determined as either a PP or noise. A preamble is declared when at less $N_{\mathrm{th}}$ elements of $\check{\boldsymbol{Z}}_m[t] $ are decided as PPs, where $N_{\mathrm{thrh}}$ is a predetermined threshold number of PPs out of $N$  to declare that $\check{\boldsymbol{Z}}_m[t] $ contains (part or whole) the preamble.

However, this approach does not take into account the statistical dependency of the consecutive PPs in the case of block fading channel, when the channel are assumed to be constant over a coherence time. It also overlooks the consecutive structure of a preamble, and the possibility of other error events caused by interference. In the remaining of this section, we study the likelihood of a preamble, which takes into account  the preamble structure and the statistical dependency between PPs, and the likelihoods of other error events that can corrupt the preamble detection in multiuser LoRa networks to derive a ML-based { non-coherent} preamble detector. Specifically, the likelihood function of $Z_{n,u}$ will be derived for the following cases: preamble, resembled preamble caused by joint-EDs, and noise plus interference.
%	\begin{itemize}
	%		\item Noise
	%		\item Preamble
	%		\item Resembled preamble
	%		\item Interference
	%	\end{itemize}
%		To avoid this effect, the simple solution is to assign the preambles  to different EDs such that the distance $\Delta_m=\mathrm{mod}\left(\left\vert\kappa_{m,1}-\kappa_{m,2}\right\vert,M/2\right)$   are unique for each and every ED in the network. Since the LoRa chirp set is periodic with the period of $M$ sample, this condition is generalized as
%		\begin{equation}
	%			\Delta_m\neq\Delta_{m^{'}},\forall m,m^{'}.
	%		\end{equation} 
\subsection{Likelihood of noise}
To determine when a preamble is present, it is also important to know when there is not. Therefore, first the likelihood that $Z_{n,u}$ is noise must be evaluated. It can be seen that, when no ED is transmitting, the $L\times 1$ signal vector $\check{\boldsymbol{V}}_{n,k}$ received on $L$ antennas on the $k$th bin at time $n$ is given as:
\begin{equation}
	\check{\boldsymbol{V}}_{n,k} = \check{\boldsymbol{W}}_{n,k},\;\forall k.
\end{equation}
As a result, the BCP signal of the $u$th at time $n$ becomes
%\begin{equation}
\begin{align}
	\label{eq-pdw-Z-noise}
	{Z}_{n,u}
	&= \frac{1}{L}\mathfrak{R}\left\{ \check{\boldsymbol{V}}_{n,\kappa_{u,1}}^{\Hi}\check{\boldsymbol{V}}_{n,\kappa_{u,2}}\right\} =  \frac{1}{L}\mathfrak{R}\left\{	\check{\boldsymbol{W}}_{n,\kappa_{u,1}}^{\Hi}\check{\boldsymbol{W}}_{n,\kappa_{u,2}}\right\}\nn\\
	&=\frac{1}{L}\sum_{\ell=1}^L\mathfrak{R}\left\{\left(\tilde{W}_{n,\ell}[\kappa_{u,1}]\right)^* \tilde{W}_{n,\ell}[\kappa_{u,2}]\right\}.
	%\xrightarrow[M\rightarrow\infty]{\mathrm{a.s}}\frac{M}{2},\nn
\end{align}
%\end{equation}
To derive the likelihood function of ${Z}_{n,u}$ being noise,  its statistic must be examined. However, the exact distribution of ${Z}_{n,u}$, which is the real part of the product of two complex-normal distributed random vectors, is rather sophisticated. Thus, instead of finding the exact distribution for ${Z}_{n,u}$ in this case, an approximated Gaussian distribution is derived as in Lemma \ref{lem1} following the central limit theorem \cite{Billingsley2005}.

\begin{lemma}\label{lem1}
	When the BCP signal of the $u$th ED at time $n$ contains only noise, the distribution of ${Z}_{n,u}$ can be approximated as
	$
	{Z}_{n,u}\sim\mathcal{N}(0,\sigma_{\mathrm{n}}^2),
	$
	%%%%%%%%%%%%
	where $\sigma_{\mathrm{n}}=\sigma^4/2L$.
\end{lemma}
\textit{Proof:} Please refer to Appendix \ref{appx-1}. $\hfill\square$

%From Lemma \ref{lem1}, the probability distribution function (pdf) that all $N$ elements of $\boldsymbol{Z}_{n,u}$ contain only noise can be calculated as in Corollary \ref{col1}.
%\begin{corollary}\label{col1}
%	The pdf of $\boldsymbol{Z}_{n,u}$ given that all $N$ elements are noise is:
%	\begin{equation}
	%		f(\boldsymbol{Z}_{n,u}|\mathrm{noise}) =\left(\frac{1}{\sqrt{2\pi\sigma^2_{\mathrm{n}}}}\right)^N\exp\left\{-\frac{\left\Vert\boldsymbol{Z}_{n,u}\right\Vert^2}{2\sigma^2_{\mathrm{n}}}
	%		\right\}
	%	\end{equation}
%\end{corollary}
%\textit{Proof:} The proof is straightforward as a result of Lemma \ref{lem1} and due to the fact that ${Z}_{n,u}$ and ${Z}_{n-M,u}$ are statistically independent. $\hfill\blacksquare$

From Lemma \ref{lem1}, it can be easily seen that the likelihood function that all $N$ elements of $\boldsymbol{Z}_{n,u}$ are noise is:
\begin{align}\label{eqLLHn}
	&\mathcal{L}^{\mathrm{(n)}}(\boldsymbol{Z}_{n,u}) = f(\boldsymbol{Z}_{n,u}|\mathrm{noise})=
	\frac{\exp\left\{-\frac{\left\Vert\boldsymbol{Z}_{n,u}\right\Vert^2}{2\sigma^2_{\mathrm{n}}}
		\right\}}{\left(\sqrt{2\pi\sigma^2_{\mathrm{n}}}\right)^N},
\end{align}
where $f(\boldsymbol{Z}_{n,u}|\mathrm{noise})$ is the probability distribution function (pdf) that all $N$ elements of $\boldsymbol{Z}_{n,u}$ contain only noise.
\subsection{Likelihood of true preamble from a single ED}
Similar to the  case of noise, when  ${Z}_{n,u}$  is a PP, it is also the sum of $L$ i.i.d. random variables. Likewise, in this case, the distribution of ${Z}_{n,u}$  can be approximated  in Lemma \ref{lem2} following the central limit theorem \cite{Billingsley2005}.
\begin{lemma}\label{lem2}
	When the BCP signal of the $u$th ED at time $n$ is a PP, the distribution of ${Z}_{n,u}$ can be approximated as a Gaussian random variable
	%%%%%%%%%%%%%
	%\begin{equation}\label{eq-dis-PP}
	${Z}_{n,u}\sim\mathcal{N}\left(\frac{M}{2},\sigma_{\mathrm{PP}}^2\right),$
	%	\end{equation}
%%%%%%%%%%%%
where $\sigma_{\mathrm{PP}}^2=M^2/4L+M\sigma^2/2L+\sigma^4/2L$.
\end{lemma}
\textit{Proof:} Please refer to Appendix \ref{appx-2}. $\hfill\square$

From Lemma \ref{lem1} and \ref{lem2}, one can easily derive a threshold to determine whether ${Z}_{n,u}$ is noise or a PP. Then, conventionally, by counting if the number of PPs in $\boldsymbol{Z}_{n,u}$ exceed a predetermined number, the GW can declare if a preamble is present. However, this decision rule overlooks two important properties of the LoRa preamble, which can be utilized to enhance the preamble detection performance:
\begin{itemize}
\item The consecutive structure of LoRa preamble: since the LoRa preamble arrives as $N$ consecutive identical symbols, if $p$ elements of $\boldsymbol{Z}_{n,u}$ are PPs ($p\leq N$), their positions in the vector   $\boldsymbol{Z}_{n,u}$ are not random, but must be either the first $p$ or the last $p$ elements of $\boldsymbol{Z}_{n,u}$. The last $p$ elements of $\boldsymbol{Z}_{n,u}$ being PPs means that the first $p$ symbols of the preamble have arrived. Whereas, the first $p$ elements of $\boldsymbol{Z}_{n,u}$ being PPs means that the preamble has been over and  its last $p$ symbols are leaving  $\boldsymbol{Z}_{n,u}$.
\item The statistical dependency among the PPs: from \eqref{eq-pdw-Z-PP}, it can be seen that the third term ${M}\Vert\boldsymbol{h}_u\Vert^2/{2L}$ of $Z_{n,u}$ does not depend on $n$. It appears in every PP element of $\boldsymbol{Z}_{n,u}$ and results in the statistical dependency of these PPs. By establishing the joint distribution of the PPs in $\boldsymbol{Z}_{n,u}$, a more precise likelihood function for the preamble can be achieved rather than examining the PPs individually as in conventional LoRa preamble detection.
\end{itemize}

Based on Lemma \ref{lem1} and \ref{lem2}, the distribution of $\boldsymbol{Z}_{n,u}$ given that $p$ out of its $N$ elements are PPs is given in Corollary \ref{col2}.
\begin{corollary}\label{col2} The distribution of $\boldsymbol{Z}_{n,u}$ given that the first $p$    elements are PPs   can be approximated as a multivariate Gaussian distribution with pdf:
% where the means are denoted as $\boldsymbol{\mu}_p= \left[\mu_{p,1},\mu_{p,2}\dots \mu_{p,N}\right]^T$:
% and $\tilde{\boldsymbol{\mu}}_p= \left[\tilde{\mu}_{p,1},\tilde{\mu}_{p,2}\dots \tilde{\mu}_{p,N}\right]^T$ with:
%\begin{figure*}[t!]
\begin{equation}\label{eq.LLfirstPP}
	\begin{split}
		&f(\boldsymbol{Z}_{n,u}|\mathrm{first }\;p\;\mathrm{ PPs}) 
		%= f(\text{first $p$ samples of $\boldsymbol{Z}_m[n]$ are PPs}) + f(\text{last $p$ samples of $\boldsymbol{Z}_m[n]$ are PPs})\\
		=\frac{\exp\left(-\frac{\left(\boldsymbol{Z}_{n,u} -\boldsymbol{\mu}_p\right)^{\Hi}\boldsymbol{\Sigma}_p^{-1}\left(\boldsymbol{Z}_{n,u} -\boldsymbol{\mu}_p\right)}{2}\right)}{(2\pi)^{N/2}\sqrt{\mathrm{det}(\boldsymbol{\Sigma}_p)}},
		%+\frac{1}{(2\pi)^N\sqrt{\mathrm{det}(\tilde{\boldsymbol{\Sigma}}_p)}}\exp\left(-\frac{\left(\boldsymbol{Z}_m[n]-\tilde{\boldsymbol{\mu}}_p\right)^{H}\tilde{\boldsymbol{\Sigma}}_p^{-1}\left(\boldsymbol{Z}_m[n]-\tilde{\boldsymbol{\mu}}_p\right)}{2}\right)
	\end{split}
\end{equation}
%\end{figure*}
%\begin{equation}
%	\begin{split}
	%	&f(\text{first $k$ samples of $\boldsymbol{Z}_m[n]$ are peaks}) =\\
	%	& \frac{1}{(2\pi)^N\sqrt{\mathrm{det}(\boldsymbol{\Sigma})}}\exp\left(-\frac{\left(\boldsymbol{Z}_m[n]-\boldsymbol{\mu}\right)^{H}\boldsymbol{\Sigma}^{-1}\left(\boldsymbol{Z}_m[n]-\boldsymbol{\mu}\right)}{2}\right)
	%	\end{split}
%\end{equation}
where $\boldsymbol{\mu}_p$ is the mean vector with the element $\mu_{p,i}=\frac{M}{2}$ for $i\leq p$ or $\mu_{p,i}=0$, otherwise,
%		\begin{equation}\label{eq.meanPPfirst}
	%			\mu_{p,i} = \begin{cases}
		%				\frac{M}{2}, \quad i\leq p\\
		%				0, \quad\text{otherwise}
		%			\end{cases}
	%		\end{equation}
and the elements of the covariance matrix $\boldsymbol{\Sigma}_p$ are
\begin{equation}
	{\Sigma}_{p,i,j} = \begin{cases}
		\sigma_{\mathrm{PP}}^2, \quad i =j\leq p \quad \\
		\sigma_{\mathrm{n}}^2, \quad i = j  > p\\
		\frac{M^2}{4L}, \quad i< j \leq p\\
		\frac{M^2}{4L}, \quad j< i \leq p\\
		0,\quad\text{otherwise}.
	\end{cases}
\end{equation}
Likewise, the pdf of $\boldsymbol{Z}_{n,u}$ when the last $p$ elements are PPs $f(\boldsymbol{Z}_{n,u}|\mathrm{last }\;p\;\mathrm{ PPs})$ is similar to \eqref{eq.LLfirstPP}, but with  the mean vector ${\boldsymbol{\mu}}_p$ replaced by $\tilde{\boldsymbol{\mu}}_p$, where $\tilde{\mu}_{p,i}=\mu_{p,N-i+1}$, and  the covariance matrix $\boldsymbol{\Sigma}_p$ replaced by $\tilde{\boldsymbol{\Sigma}}_p$, { whose element $\tilde{\Sigma}_{p,i,j}={\Sigma}_{p,N-i+1,N-j+1}$.}
%		Whereas the pdf of $\boldsymbol{Z}_{n,u}$ when the last $p$ elements are PPs  is:
%		
%		\begin{equation}\label{eq.LLlastPP}
	%			\begin{split}
		%				&f(\boldsymbol{Z}_{n,u}|\mathrm{last }\;p\;\mathrm{ PPs}) 
		%				%= f(\text{first $p$ samples of $\boldsymbol{Z}_m[n]$ are PPs}) + f(\text{last $p$ samples of $\boldsymbol{Z}_m[n]$ are PPs})\\
		%				=\frac{\exp\left(-\frac{\left(\boldsymbol{Z}_{n,u} -\tilde{\boldsymbol{\mu}}_p\right)^{\Hi}\tilde{\boldsymbol{\Sigma}}_p^{-1}\left(\boldsymbol{Z}_{n,u} -\tilde{\boldsymbol{\mu}}_p\right)}{2}\right)}{(2\pi)^{N/2}\sqrt{\mathrm{det}(\tilde{\boldsymbol{\Sigma}}_p)}}
		%				%+\frac{1}{(2\pi)^N\sqrt{\mathrm{det}(\tilde{\boldsymbol{\Sigma}}_p)}}\exp\left(-\frac{\left(\boldsymbol{Z}_m[n]-\tilde{\boldsymbol{\mu}}_p\right)^{H}\tilde{\boldsymbol{\Sigma}}_p^{-1}\left(\boldsymbol{Z}_m[n]-\tilde{\boldsymbol{\mu}}_p\right)}{2}\right)
		%			\end{split}
	%		\end{equation}
%		where the mean vector $\tilde{\boldsymbol{\mu}}_p$ has the element $\tilde{\mu}_{p,i}=\mu_{p,N-i+1}$
%%		
%%		\begin{equation}
	%%			\tilde{\mu}_{p,i} = \begin{cases}
		%%				\frac{M}{2}, \quad i\geq N-p+1\\
		%%				0, \quad\text{otherwise}
		%%			\end{cases}
	%%		\end{equation}
%		and  the covariance matrix $\tilde{\boldsymbol{\Sigma}}_p=\boldsymbol{\Sigma}_p^{\tilde{\T}}$.
%		\begin{equation}
	%			\tilde{\Sigma}_{p,i,j} = \begin{cases}
		%				\sigma_{\mathrm{PP}}^2, \quad i =j\geq N-p+1 \quad \\
		%				\frac{\sigma^4}{2L}, \quad i = j  < N-p+1\\
		%				\frac{M^2}{4L}, \quad i> j \geq N-p+1\\
		%				\frac{M^2}{4L}, \quad j> i \geq N-p+1\\
		%				0,\quad\text{otherwise}
		%			\end{cases}
	%		\end{equation}
	\end{corollary}
	\textit{Proof:} Please refer to Appendix \ref{appx-3}. $\hfill\square$
	
	From Corollary \ref{col2}, the likelihood function that a part of the preamble of the $u$th ED is present in $\boldsymbol{Z}_{n,u}$ is given as:
	\begin{align}\label{eq.LLHpp}
\mathcal{L}^{\mathrm{(p)}}(\boldsymbol{Z}_{n,u})&=\sum_{p=N_{\mathrm{thr}}}^{N}f(\boldsymbol{Z}_{n,u}|\mathrm{first }\;p\;\mathrm{ PPs})\nn\\ &+\sum_{p=N_{\mathrm{thr}}}^{N-1}f(\boldsymbol{Z}_{n,u}|\mathrm{last }\;p\;\mathrm{ PPs}). 
\end{align}
Note that the two sums in the above equation only run from $N=N_{\mathrm{thr}}$, which is the minimum number of PPs required in $\boldsymbol{Z}_{n,u}$ to declare a preamble. This is to lowers the chance that noise and interference can be mistaken to a true preamble when the required number of PPs is too low. A higher value of $N_{\mathrm{thr}}$ means a stricter condition to declare a preamble, which lower the probability of mistaking noise and interference to true preambles, but can also increase the probability of missing a presence under the influence of fading, noise and interference. By contrast, a lower value of $N_{\mathrm{thr}}$  results in the opposite.
In the simulation results, the changes in preamble detection performance when varying $N_{\mathrm{thr}}$ will be observed.

In addition, the second sum only runs from $N_{\mathrm{thr}}$ to $N-1$, since with $p=N$, $f(\boldsymbol{Z}_{n,u}|\mathrm{first }\;p\;\mathrm{ PPs})$ and $f(\boldsymbol{Z}_{n,u}|\mathrm{last }\;p\;\mathrm{ PPs})$ are the same, which is when all $N$ elements of $\boldsymbol{Z}_{n,u}$ are PPs. Thus, the value of $f(\boldsymbol{Z}_{n,u}|\mathrm{last }\;i\;\mathrm{ PPs})$ for $p=N$ is discarded to avoid duplicate.
\subsection{Likelihood of resembled preamble}
Similar to the case of true preambles,  resembled preambles also have the consecutive-identical-symbol structure and the dependency between resembled PPs (RPPs). Thus, the distribution of resembled PPs must be derived similar to a true preamble. First, the distribution of $Z_{n,u}$ given that it is a RPP can be found as in Lemma \ref{lem3} below.

\begin{lemma}\label{lem3}
When the BCP signal of the $u$th ED at time $n$ is a RPP, the distribution of ${Z}_{n,u}$ can be approximated as
%%%%%%%%%%%%%
%\begin{equation}\label{eq-dis-RPP}
${Z}_{n,u}\sim\mathcal{N}\left(0,\sigma_{\mathrm{R}}^2\right),$
%\end{equation}
%%%%%%%%%%%%
where $\sigma_{\mathrm{R}}^2=M^2/8L+M\sigma^2/2L+\sigma^4/2L$.
\end{lemma}

\textit{Proof:} The proof of this lemma is straightforward and can be performed similar to that of Lemma \ref{lem2}. $\hfill\blacksquare$ 
%Please refer to Appendix \ref{appx-4}. $\hfill\square$

Similar to the case of true preamble, considering the consecutive-identical-symbol structure of preamble resemblance, the distribution of $\boldsymbol{Z}_{n,u}$ given that $p$ out of $N$ of its elements are RPPs can be derived as in the following Corollary.

\begin{corollary}\label{col3} The distribution of $\boldsymbol{Z}_{n,u}$ given that the first $p$    elements are RPPs   can be approximated as a multivariate Gaussian distribution with pdf:
% where the means are denoted as $\boldsymbol{\mu}_p= \left[\mu_{p,1},\mu_{p,2}\dots \mu_{p,N}\right]^T$:
% and $\tilde{\boldsymbol{\mu}}_p= \left[\tilde{\mu}_{p,1},\tilde{\mu}_{p,2}\dots \tilde{\mu}_{p,N}\right]^T$ with:
%\begin{figure*}[t!]
\begin{equation}\label{eq.LLfirstRPP}
	\begin{split}
		&f(\boldsymbol{Z}_{n,u}|\mathrm{first }\;p\;\mathrm{ RPPs}) 
		%= f(\text{first $p$ samples of $\boldsymbol{Z}_m[n]$ are PPs}) + f(\text{last $p$ samples of $\boldsymbol{Z}_m[n]$ are PPs})\\
		=\frac{\exp\left(-\frac{\boldsymbol{Z}_{n,u}^{\Hi}\boldsymbol{\Psi}_p^{-1}\boldsymbol{Z}_{n,u}}{2}\right)}{(2\pi)^{N/2}\sqrt{\mathrm{det}(\boldsymbol{\Psi}_p)}},
		%+\frac{1}{(2\pi)^N\sqrt{\mathrm{det}(\tilde{\boldsymbol{\Sigma}}_p)}}\exp\left(-\frac{\left(\boldsymbol{Z}_m[n]-\tilde{\boldsymbol{\mu}}_p\right)^{H}\tilde{\boldsymbol{\Sigma}}_p^{-1}\left(\boldsymbol{Z}_m[n]-\tilde{\boldsymbol{\mu}}_p\right)}{2}\right)
	\end{split}
\end{equation}
%\end{figure*}
%\begin{equation}
%	\begin{split}
	%	&f(\text{first $k$ samples of $\boldsymbol{Z}_m[n]$ are peaks}) =\\
	%	& \frac{1}{(2\pi)^N\sqrt{\mathrm{det}(\boldsymbol{\Sigma})}}\exp\left(-\frac{\left(\boldsymbol{Z}_m[n]-\boldsymbol{\mu}\right)^{H}\boldsymbol{\Sigma}^{-1}\left(\boldsymbol{Z}_m[n]-\boldsymbol{\mu}\right)}{2}\right)
	%	\end{split}
%\end{equation}
where the elements of the covariance matrix $\boldsymbol{\Psi}_p$ are
\begin{equation}
	{\Psi}_{p,i,j} = \begin{cases}
		\sigma_{\mathrm{R}}^2, \quad i =j\leq p \quad \\
		\sigma_{\mathrm{n}}^2, \quad i = j  > p\\
		\frac{M^2}{8L}, \quad i< j \leq p\\
		\frac{M^2}{8L}, \quad j< i \leq p\\
		0,\quad\text{otherwise}.
	\end{cases}
\end{equation}
Likewise, the pdf of $\boldsymbol{Z}_{n,u}$ when the last $p$ elements are RPPs $f(\boldsymbol{Z}_{n,u}|\mathrm{last }\;p\;\mathrm{ RPPs})$ is similar to \eqref{eq.LLfirstRPP}, but with   the covariance matrix $\boldsymbol{\Psi}_p$ replaced by $\tilde{\boldsymbol{\Psi}}_p$, { whose element  $\tilde{\Psi}_{p,i,j}={\Psi}_{p,N-i+1,N-j+1}$}.

%		\begin{equation}\label{eq.LLlastRPP}
	%			\begin{split}
		%				&f(\boldsymbol{Z}_{n,u}|\mathrm{last }\;p\;\mathrm{ RPPs}) 
		%				%= f(\text{first $p$ samples of $\boldsymbol{Z}_m[n]$ are PPs}) + f(\text{last $p$ samples of $\boldsymbol{Z}_m[n]$ are PPs})\\
		%				=\frac{\exp\left(-\frac{\boldsymbol{Z}_{n,u}^{\Hi}\tilde{\boldsymbol{\Psi}}_p^{-1}\boldsymbol{Z}_{n,u}}{2}\right)}{(2\pi)^{N/2}\sqrt{\mathrm{det}(\tilde{\boldsymbol{\Psi}}_p)}}.
		%				%+\frac{1}{(2\pi)^N\sqrt{\mathrm{det}(\tilde{\boldsymbol{\Sigma}}_p)}}\exp\left(-\frac{\left(\boldsymbol{Z}_m[n]-\tilde{\boldsymbol{\mu}}_p\right)^{H}\tilde{\boldsymbol{\Sigma}}_p^{-1}\left(\boldsymbol{Z}_m[n]-\tilde{\boldsymbol{\mu}}_p\right)}{2}\right)
		%			\end{split}
	%		\end{equation}
%		where  the covariance matrix $\tilde{\boldsymbol{\Psi}}_p=\boldsymbol{\Psi}_p^{\tilde{\T}}$.
%		\begin{equation}
	%			\tilde{\Psi}_{p,i,j} = \begin{cases}
		%				\sigma_{\mathrm{R}}^2, \quad i =j\geq N-p+1 \quad \\
		%				\frac{\sigma^4}{2L}, \quad i = j  < N-p+1\\
		%				\frac{M^2}{8L}, \quad i> j \geq N-p+1\\
		%				\frac{M^2}{8L}, \quad j> i \geq N-p+1\\
		%				0,\quad\text{otherwise}
		%			\end{cases}
	%		\end{equation}
	\end{corollary}
	\textit{Proof:} The proof of this corollary is straightforward and can be performed similar to that of Corollary \ref{col2}. $\hfill\blacksquare$
	
	From Lemma \ref{lem3}, the likelihood that $\boldsymbol{Z}_{n,u}$ contains a part of a resembled preamble can be calculated as:
	\begin{equation}\label{eqLLr}
\begin{split}
	\mathcal{L}^{\mathrm{(r)}}(\boldsymbol{Z}_{n,u})&=\sum_{p=1}^{N}f(\boldsymbol{Z}_{n,u}|\mathrm{first }\;p\;\mathrm{ RPPs})\\ &+\sum_{p=1}^{N-1}f(\boldsymbol{Z}_{n,u}|\mathrm{last }\;p\;\mathrm{ RPPs}).
\end{split}
\end{equation}
Similar to the case of true preamble, the second sum in \eqref{eqLLr} only runs from $p=1$ to $N-1$ to avoid duplicate.

\subsection{Likelihood of other interference}
In order for the $\tilde{u}$th ED to interfere with the BCP signal $Z_{n,u}$ of the $u$th ED, its signal  $\bar{\boldsymbol{x}}_{n,\tilde{u}}$ must simultaneously occupy the $\kappa_{u,1}$th and $\kappa_{u,2}$th bin.
According to the analysis in Section~\ref{Sec:preamble}-B, this cannot be caused by the preamble of the $\tilde{u}$th ED, thanks to the preamble design and assignment. However, this can happen when  $\bar{\boldsymbol{x}}_{n,\tilde{u}}$ contains the fragments of two different chirps, when the detection window is moving from one symbol to the next one, as analyzed in Section~\ref{Sec:preamble}-B. According to Fig~\ref{fig-bin}, the amplitudes of the bins when $\bar{\boldsymbol{x}}_{n,\tilde{u}}$ contains the fragments of two different chirps are of small values. Thus, this type of interference can be treated as part of the noise.

The interference can also be jointly caused by multiple EDs, the same way  resembled preambles are created. However, thanks to the asymptotic orthogonality among different ED's channels, this type of interference converges to zero when the number of antennas tends to infinity \cite{Marzetta2016}. Since this type of interference is also of zero mean and small amplitudes, it can also be treated as noise.

Thus, in this paper, we do not derive the  distribution for interference. The reason is that the interaction between the signals of multiple EDs (causes joint-interference from multiple EDs), and different fragments of chirps (causes interference from a single ED) are extremely complicated, and is mostly intractable. In addition, as discussed above, the characteristics of interference can be reasonably treated as noise. This is also a common assumption in existing works in preamble detection \cite{Kang2022,Kang2024,Kang2023}.

\subsection{ML-based non-coherent double-chirp preamble detection}
From the above likelihood functions,  the preamble detection rule for the $u$th ED can be established as follows:
%%%%%%%%%%%%%%%%
\begin{equation}\label{eq.MLPD}
\mathcal{L}^{\mathrm{(p)}}(\boldsymbol{Z}_{n,u}) \overset{\text{preamble}}{\underset{\text{not preamble}}{\gtrless}} \mathcal{L}^{\mathrm{(n)}}(\boldsymbol{Z}_{n,u}) + \mathcal{L}^{\mathrm{(r)}}(\boldsymbol{Z}_{n,u}). %+ \mathcal{L}^{\mathrm{(i)}}(\boldsymbol{Z}_{n,u})
\end{equation}
%%%%%%%%%%%%%%%%%%
Once this condition is satisfied, the GW will determine that the preamble of the $u$th ED is present. The proposed ML-based preamble detection method for multiuser massive MIMO LoRaWAN is summarized in the following Algorithm \ref{al:al2}.

%	Suppose that at least $N_{\mathrm{thr}}$ PPs are required to be present in  $\boldsymbol{Z}_{n,u}$  to declare a preamble. When $p>N_{\mathrm{thr}}$ PPs are present, it is even more likely that the ML-based detector in \eqref{eq.MLPD} will also declare that a preamble is present. Thus, instead of immediately declaring the presence of a preamble when \eqref{eq.MLPD} is satisfied, the GW can  wait for $N_{\mathrm{rep}}$  consecutively repeated preamble alerts, distanced by $M$ samples, to make sure that it is truly preamble, not noise and interference. If the preamble alerts are repeated less than $N_{\mathrm{rep}}$ times, or the distance between different alerts are not $M$ samples, the GW will not declare a preamble, and reset the counter for the preamble alert repetitions to zero. This algorithm is summarized in Algorithm~\ref{al:al2}.
%	
%	{  Comments: According to the simulation results, this algorithm does not make any significant different to the performance, so it should be removed.}

\begin{algorithm}[tb]
\caption{ML-based double-chirp  preamble detection}
{\small
	\begin{algorithmic}[1]\label{al:al2}
		\REQUIRE Number of EDs, SF, preamble assignment.
		\STATE $AllDetected=0$, $counter[u]=0,\forall u$, $last[u]=0$
		%\STATE  $\mathcal{T}_u=\emptyset,\forall u$;
		%\STATE $\alpha_u=0$, $\forall u$;
		%\STATE Initiating LLR storage $\boldsymbol{\ell}_m^{\mathrm{(save)}} = [\emptyset], \forall m$
		\WHILE{$AllDetected=0$} 
		\STATE Performing dechirped-FFT to obtain $\check{\boldsymbol{V}}_{n,\ell},\forall\ell$;% and %$\bar{\boldsymbol{V}}_{n,\ell}$;
		\FOR{$u=1$ to $N_u$}
		
		\STATE Calculating $	{\boldsymbol{Z}}_{n,u}$, and  likelihood functions    \eqref{eqLLHn}, \eqref{eq.LLHpp}, \eqref{eqLLr};
		
		\IF{$\mathcal{L}^{\mathrm{(p)}}(\boldsymbol{Z}_{n,u}) > \mathcal{L}^{\mathrm{(n)}}(\boldsymbol{Z}_{n,u}) + \mathcal{L}^{\mathrm{(r)}}(\boldsymbol{Z}_{n,u})$}
		%\STATE $last[u]= n$;
		%\STATE $\alpha_u= \mathrm{size}(\textbf{r})$;
		%				\IF {$(\mathrm{mod}(n-last[u],M)=0)$}
		%				\STATE $count[u] = count[u] +1$;
		%				\STATE $last[u]= n$;
		%				\IF{$count[u]\geq N_{\mathrm{rep}}$}
		\STATE  Preamble detected for the $u$th ED;
		%\STATE $\alpha_u=0$;
		%\ELSE 
		%
		%
		%				\ENDIF
		%				\ELSE
		%				\STATE $count[u] = 1$;
		%				\STATE $last[u]= n$;
		%				%\STATE $last[u]=n$
		%				\ENDIF
		%\STATE $\alpha_u = \alpha_u+1$;
		%\STATE $\textbf{r}(\alpha_u).\tau = n$;
		%\STATE Calculate $\textbf{r}(\alpha_u).\mathrm{LLH}$ as in \eqref{eq72};
		%\ELSE
		%\IF{($n = \textbf{r}(\alpha).\tau +M$) \& ($\alpha > 1$)}
		%\STATE  $i_0 = \underset{i}{\mathrm{argmax}}\left\{\textbf{r}_u(i).\mathrm{LLR} \right\}$;
		%\STATE $	\mathrm{ToA}_u = \textbf{r}_u(i_0).\tau-M(N+2)$;
		%\ENDIF				
		\ENDIF
		\ENDFOR
		\IF{All preambles are detected}
		\STATE $AllDetected=1$;
		\ENDIF
		\STATE $n=n+1$;
		%					\STATE Solve \eqref{opt2} or calculate $\boldsymbol{\varphi}_i$ from \eqref{eq.LS}.
		%					\STATE Correlate $\boldsymbol{\varphi}_i$ with preamble signal as in \eqref{eq.pilot2} to mitigate interference.
		%					\STATE Obtain estimated channel $\hat{\boldsymbol{h}}_i$ with \eqref{eq.MMSE}.
		%					\STATE Perform signal combining with MRC and DFT receiver following \eqref{eq_pilotx}, \eqref{eq3b} and \eqref{fft}. 
		%					\STATE Detect data payload of the $i$th ED's packet as in \eqref{eq:non-coh}.
		\ENDWHILE
		%			\STATE Initially detect $\tau_c$ symbols $\hat{\boldsymbol{m}}=[\hat{m}_1,\ldots,\hat{m}_{\tau_c}]$ with non-coherent detection.
		%			\STATE Estimate $\hat{h}_{\ell,i}$ corresponding to $\tau_c$ detected LoRa symbols ($i=1,2,\dots,\tau_c$).
		%			\STATE Obtain the average channel estimate $\hat{h}_{\ell}^{(\mathrm{ave})}$ as in \eqref{eq23}.
		%			\STATE Save current decision as $\hat{\boldsymbol{m}}_{\mathrm{prev}}=\hat{\boldsymbol{m}}$.
		%			\STATE $\mathrm{flag}=1$; $\mathrm{count=0;}$
		%			\WHILE{$\mathrm{flag}=1$ $\&$ $\mathrm{count}\leq N_{\mathrm{max}}$}
		%			\STATE $\mathrm{count}=\mathrm{count}+1;$
		%			\STATE For the $i$th symbol, combine $V_{\ell,i}[k]$ with $\hat{h}_{\ell}^{(\mathrm{ave})}\exp\left(j\Psi_{k}\right)$ as in \eqref{eq:dev-semi-coh} and perform a new detection of the $i$th symbol $\hat{m}_i$ as in \eqref{eq:max-semi-coh}, $\forall i=1,2,\dots,\tau_c$.
		%			\STATE Update $\hat{h}_{\ell}^{(\mathrm{ave})}$ using newly detected symbols $\hat{\boldsymbol{m}}$.
		%			\IF {$\hat{\boldsymbol{m}}_{\mathrm{prev}}\neq\hat{\boldsymbol{m}}$}
		%			\STATE $\mathrm{flag}=1$
		%			\ELSE
		%			\STATE $\mathrm{flag}=0$
		%			\ENDIF
		%			\STATE $\hat{\boldsymbol{m}}_{\mathrm{prev}}=\hat{\boldsymbol{m}}$
		%			\ENDWHILE
		\RETURN All detected preamble.
	\end{algorithmic}
}
\end{algorithm}

\section{Complexity Reduction}\label{Sec:com}
Since the detection for double-chirp preamble design requires computation \emph{every sample} (rather than every LoRa symbol as the methods in\cite{Ghanaatian2019,Tang2019,Edward2019,bernier2020}), the complexity can be an issue with  practical implementation. Thus, in this section, we provide the complexity analysis for the detection of the proposed double-chirp preamble, and how to reduce that for practical implementation.

With the fast Fourier Transform (FFT), the complexity of performing the dechirped-FFT on { all} $L$ antennas requires $\mathcal{O}(LM\mathrm{log}_2M)$. Then, to obtain the preamble BCP signal in \eqref{eq-pdw-Z-PP} of all $N_u$ EDs, the complexity is $\mathcal{O}(N_uL)$. The calculation of the likelihood functions \eqref{eqLLHn}, \eqref{eq.LLHpp}, \eqref{eqLLr} for all $N_u$ EDs requires $\mathcal{O}(N_u N^3)$. As a result, the overall computational complexity to simultaneously detect the preambles of $N_u$ EDs is $\mathcal{O}(LM\mathrm{log}_2M+ N_uL +N_u N^3)$ every sample. Since $M$ is significantly larger than $N_u$, $L$ and $N$, it can be seen that the most computationally demanding task  is the FFT. Thus, we propose the following technique to relax the GW from having to compute the FFT for every sample.

Suppose that the bin signal vectors ${\boldsymbol{V}}_{n,\ell}$ on all $M$ bins in the frequency domain are calculated in \eqref{eq-pdw-V} for time $n$. From this equation, the bin signals corresponding to time $n+1$ are:

%%%%%%%%%%%%%%%%%%%
\begin{equation}\label{eq-pdw-V2}
{\boldsymbol{V}}_{n+1,\ell}= \mathrm{DFT}\left\{\boldsymbol{y}_{n+1,\ell}\otimes\boldsymbol{s}_0^*\right\}\in \mathbb{C}^{M\times1}.
\end{equation}
%%%%%%%%%%%%%%%%%%
%	%%%%%%%%%%%%%%%%%%%%%%%%%%%%%%%%%%
%	\begin{equation}\label{eq-pdw-V2}
%		\begin{split}
	%			\boldsymbol{V}_{\ell}^{\mathrm{(PDW)}}[n+1]&= \mathrm{DFT}\left\{\boldsymbol{d}_{\ell}[n+1]\otimes\boldsymbol{x}_0^*\right\}\in \mathbb{C}^{M\times1}.\\
	%		\end{split}
%		\end{equation}
%		%%%%%%%%%%%%%%%%%%
%%%%%%%%%%%%%%%%%
Since
%%%%%%%%%%%%%
$
\boldsymbol{y}_{n+1,\ell} = \left(\boldsymbol{y}_{n,\ell} \ll1\right) + [0,0\dots0,\delta_{n+1}]^{\T},
$
%%%%%%%%%%%%%
where  $\delta_{n+1}= y_{n+M,\ell}-y_{n+M-1,\ell}$, we have:
%%%%%%%%%%%%%%%%
%\begin{align}\label{eq30}
%%\begin{split}
%	&\boldsymbol{y}_{n+1,\ell}\otimes\boldsymbol{s}_0^*
%	= \left(\boldsymbol{y}_{n,\ell}\ll1\right)\otimes\boldsymbol{s}_0^* + [0,\dots0,\tilde{\delta}_{n+1}]^{\T}\nn\\
%	&=\left(\left(\boldsymbol{y}_{n,\ell}\otimes\left(\boldsymbol{s}_0^*\gg1\right)\right)\ll1\right)+ [0,\dots0,\tilde{\delta}_{n+1}]^{\T},
%%\end{split}
%\end{align}
%%%%%%%%%%%%%%%%%%%
%%%%%%%%%%%%%%%
\begin{align}\label{eq30}
%\begin{split}
\boldsymbol{y}_{n+1,\ell}\otimes\boldsymbol{s}_0^*
&= \left(\boldsymbol{y}_{n,\ell}\ll1\right)\otimes\boldsymbol{s}_0^* + [0,\dots0,{\delta}_{n+1}]^{\T}\otimes\boldsymbol{s}_0^*\nn\\
&=\left(\left(\boldsymbol{y}_{n,\ell}\otimes\left(\boldsymbol{s}_0^*\gg1\right)\right)\ll1\right)+ \boldsymbol{d},
%\end{split}
\end{align}
%%%%%%%%%%%%%%%%%%
with  $\boldsymbol{d}=s^{*}_{0,m}[0,\dots0,{\delta}_{n+1}]^{\T}$. Plugging \eqref{eq30} into \eqref{eq-pdw-V2} yields:
%%%%%%%%%%
\begin{align}\label{eq31}
%\begin{split}
{\boldsymbol{V}}_{n+1,\ell}  
&=\mathrm{DFT}\left\{\left(\left(\boldsymbol{y}_{n,\ell}\otimes\left(\boldsymbol{s}_0^*\gg1\right)\right)\ll1\right)+\boldsymbol{d}\right\}\nn\\
&=\mathrm{DFT}\left\{\left(\boldsymbol{y}_{n,\ell}\otimes\boldsymbol{s}_1^*\right)\ll1\right\} +	\mathrm{DFT}\left\{\boldsymbol{d}\right\}.
%\end{split}
\end{align}
%%%%%%%%%%%%%%%%
The second term of \eqref{eq31} is the DFT of a vector with only one non-zero element, which can be easily calculated in $\mathcal{O}(M)$ without using the FFT. Whereas, following the shifting properties of the DFT, the first term of \eqref{eq31} can be found as:
%%%%%%%%%%%%%%%%%
\begin{equation}\label{eq32}
\begin{split}
	&\mathrm{DFT}\left\{\left(\boldsymbol{y}_{n,\ell}\otimes\boldsymbol{s}_1^*\right)\ll1\right\}	=\boldsymbol{\alpha} \otimes \mathrm{DFT}\left\{\boldsymbol{y}_{n,\ell}\otimes\boldsymbol{s}_1^*\right\}
\end{split}
\end{equation}
%%%%%%%%%%%%%%%%%%%%
where $\boldsymbol{\alpha} = [\alpha[0],\alpha[1]\dots\alpha[M-1]]$ and $\alpha[k]=\exp\{j2\pi k/M\}$. Next, since 
%%%%%%%%%%%%%%%%%%
\begin{align}
&s_{1,m}=s_{0,m+1} =\exp\left\{j2\pi\left(\frac{ m^2}{2M}-\frac{m}{2}\right)\right\}\nn\\
&=\exp\left\{j2\pi\left(\frac{1}{2}-\frac{1}{2M}\right)\right\}\exp\left\{j2\pi \frac{m-1}{M}\right\}s_{0,m},
\end{align}
%%%%%%%%%%%%%%
we have
%%%%%%%%%%%
\begin{equation}\label{eq34}
\begin{split}
	&\mathrm{DFT}\left\{\boldsymbol{y}_{n,\ell}\otimes\boldsymbol{s}_1^*\right\} =\\
	&= \exp\left\{j2\pi\left(\frac{1}{2}-\frac{1}{2M}\right)\right\}\mathrm{DFT}\left\{\boldsymbol{y}_{n,\ell}\otimes\boldsymbol{s}_0^*\right\} \gg 1\\
	&= \exp\left\{j2\pi\left(\frac{1}{2}-\frac{1}{2M}\right)\right\}{\boldsymbol{V}}_{n,\ell}\gg1.
\end{split}
\end{equation}
%%%%%%%%%%%%%%%
Plugging \eqref{eq32}, \eqref{eq34} into \eqref{eq31}, we have:
\begin{equation}\label{eq35}
\begin{split}
	{\boldsymbol{V}}_{n+1,\ell} =\tilde{\boldsymbol{\alpha}}\otimes\left({\boldsymbol{V}}_{n,\ell}\gg1\right)+	\mathrm{DFT}\left\{\boldsymbol{d}\right\},
\end{split}
\end{equation}
where $\tilde{\boldsymbol{\alpha}}=\exp\left\{j2\pi\left(\frac{1}{2}-\frac{1}{2M}\right)\right\}\boldsymbol{\alpha}$.
This equation suggests that, when detecting preamble in every sample, only the first sample needs an FFT at time $n$. The bin signals in the frequency domain of succeeding samples  ${\boldsymbol{V}}_{n+1,\ell}$ at time $n+1$ can be calculated sequentially as \eqref{eq35} with the complexity of $\mathcal{O}(M)$, rather than $\mathcal{O}(M\mathrm{log}_2M)$ when using FFT. Consequently, the overall complexity to process one sample with the proposed preamble design is $\mathcal{O}(LM+ N_uL +N_u N^3)$, or equivalently  $\mathcal{O}(LM^2+ MN_uL + MN_u N^3)$ per symbol. 
In comparison with the conventional single-chirp preamble detection with the complexity of  $\mathcal{O}(LM\mathrm{log}_2M+ N_uL)$, the complexity of the proposed method is significantly higher. However, given the ability of multiuser detection and packet-loss avoidance to increase both the capacity and connectivity in LoRaWAN, this cost can be well justified.

\section{Simulation Results}\label{Sec:sim}

In this section, simulation results are provided to show the merits of the proposed scheme. We consider a single GW operate at 2.4 GHz with a varying number of antennas (16, 32, 64), receiving packets from multiple EDs\footnote{For this frequency, to maintain the dependency between antenna elements, a $8\times8$ antenna deployment (64 antennas) requires the size of $0.43\times0.43$ cm for the minimum spacing of half a wavelength (6.25 cm), or $0.85\times0.85$ cm for one-wavelength spacing (12.5 cm), which is practical for implementation.}. The simulation results are averaged over $10^4$ iterations. In every iteration, the packets from all EDs, the ToAs, and the channel vectors are randomly generated. Each ED is assumed to send one packet with $\mathrm{SF}=7$ and random ToA, with the payload lengths uniformly distributed from 20 to 30 symbols. The packets from different EDs are generated with the  ToA differences following an exponential distribution with the mean of one packet per one LoRa symbol duration.

\begin{figure}[t!]
\centering
\includegraphics[width=0.5\textwidth]{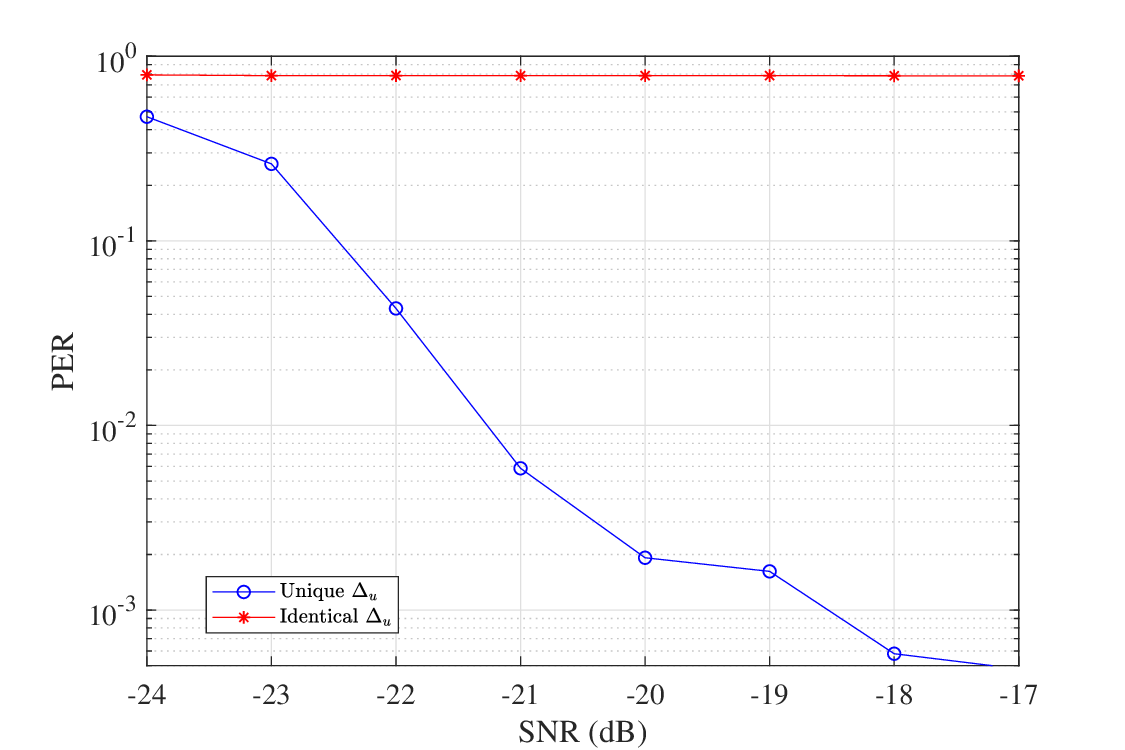}
\caption{PER with different preamble assignment ($L=32$, $N=8$, $N_{\mathrm{thr}}=4$)}.
%\label{fig.1}
\label{fig.PER4}
\end{figure}

Fig.~\ref{fig.PER4} shows the effect of preamble assignment on the {  preamble error rate (PER)} of the proposed design. When all EDs have the same $\Delta_u$, resembled preamble caused by single-ED sources will certainly occur, which results in a  very poor PER, as can be seen in the red figure. Fortunately, this can be avoided with preamble assignment by setting unique values of $\Delta_u$ for every ED, which can be seen in blue. Given the importance of preamble assignment, in the remaining figures, the simulation results are obtained when the preambles are assigned with unique values of $\Delta_u$.

\begin{figure}[t!]
\centering
\includegraphics[width=0.5\textwidth]{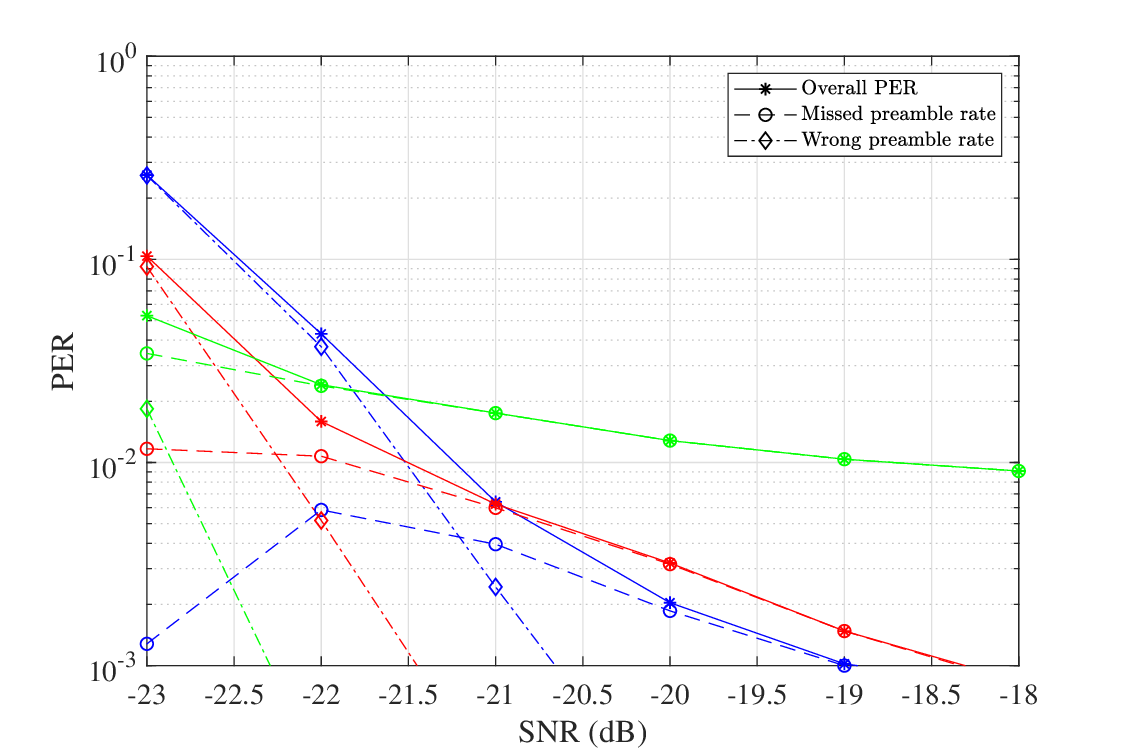}
\caption{PER with different threshold numbers of PPs ($N=8$, $L=32$, 5 EDs). Blue, red and green lines are for $N_{\mathrm{thr}}=4$, $6$ and $8$, respectively.}
%\label{fig.1}
\label{fig.PER5}
\end{figure}
Fig.~\ref{fig.PER5} shows the effect of the threshold number of PPs $N_\mathrm{thr}$ on the preamble detection performance of the proposed design. Although the overall PER is somewhat equivalent when $N_\mathrm{thr}=4$ or 6 PPs, the differences in missed preamble rate (the GW fails to detect the presence of a preamble) and wrong preamble rate (the GW mistakes noise and interference to true preamble) are significant. With the larger value of $N_{\mathrm{thr}}$, the missed preamble rate is significantly higher. As discussed { earlier}, the reason is that higher values of PPs required to declare a preamble means a stricter condition that can result in more missed preamble. This, however, will be able  to obtain a lower chance of mistaking noise and interference for a true preamble. This becomes more evident when $N_\mathrm{thr}=8$, when the wrong preamble rate is significantly lower than that of the two previous cases, but the missed preamble rate is much higher than when $N_\mathrm{thr}=4$ or 6. Given that at higher SNR range, the missed preamble rate becomes the dominant source of errors, $N_{\mathrm{thr}}=4$ appears to be a good choice for the balance between missed preamble rate and wrong-preamble rate, which results in a better overall PER. In the remaining figures, the simulation results shall be obtained with $N_{\mathrm{thr}}=4$.

\begin{figure}[t!]
\centering
\includegraphics[width=0.5\textwidth]{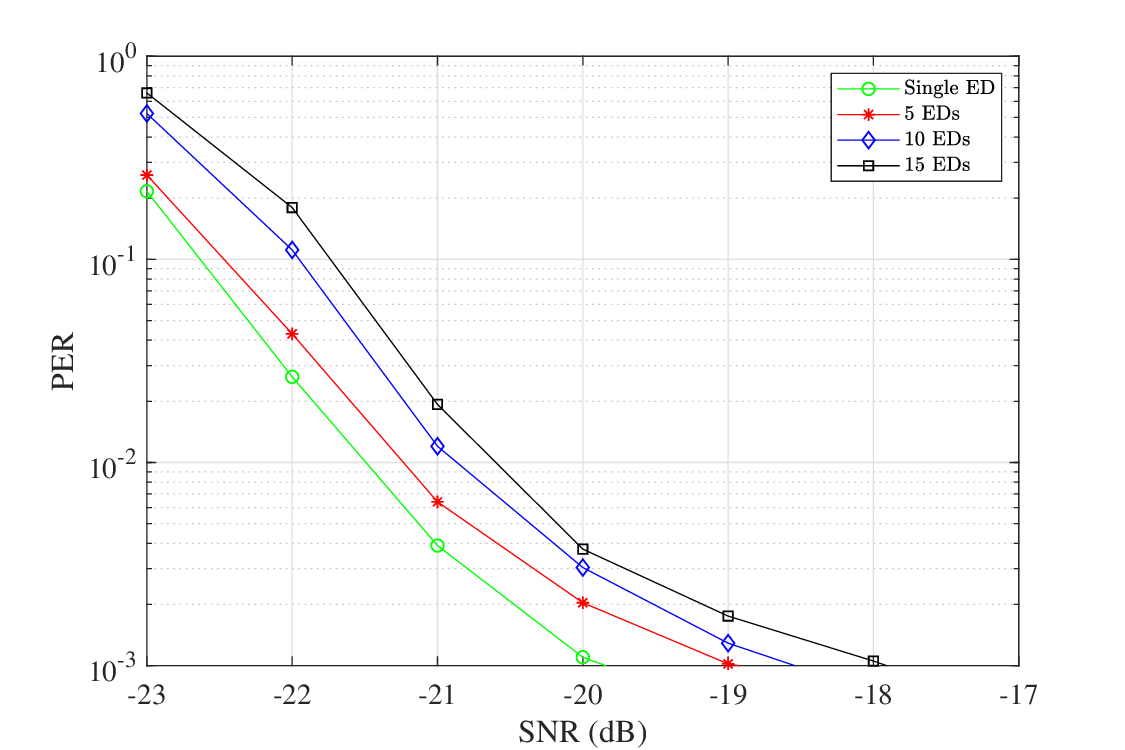}
\caption{PER with different numbers of EDs ($L=32$, $N=8$)}.
%\label{fig.1}
\label{fig.PER1}
\end{figure}

Fig.~\ref{fig.PER1} shows the PER with different numbers of EDs. An error in preamble detection is defined as when the GW fails to recognize the presence of a preamble, or mistakes noise, interference or resembled preambles as  true ones. Obviously, the PER increases when  there are more EDs concurrently transmitting packets, which results in more interference and a higher chance of resembled preamble occurring. According to the figure, at the PER of $10^{-3}$, the proposed preamble design required 2 dB more power to serve 15 EDs, as compared to the single-ED case. This appears to be a good trade-off to increase the connectivity and capacity in LoRaWAN.

\begin{figure}[t!]
\centering
\includegraphics[width=0.5\textwidth]{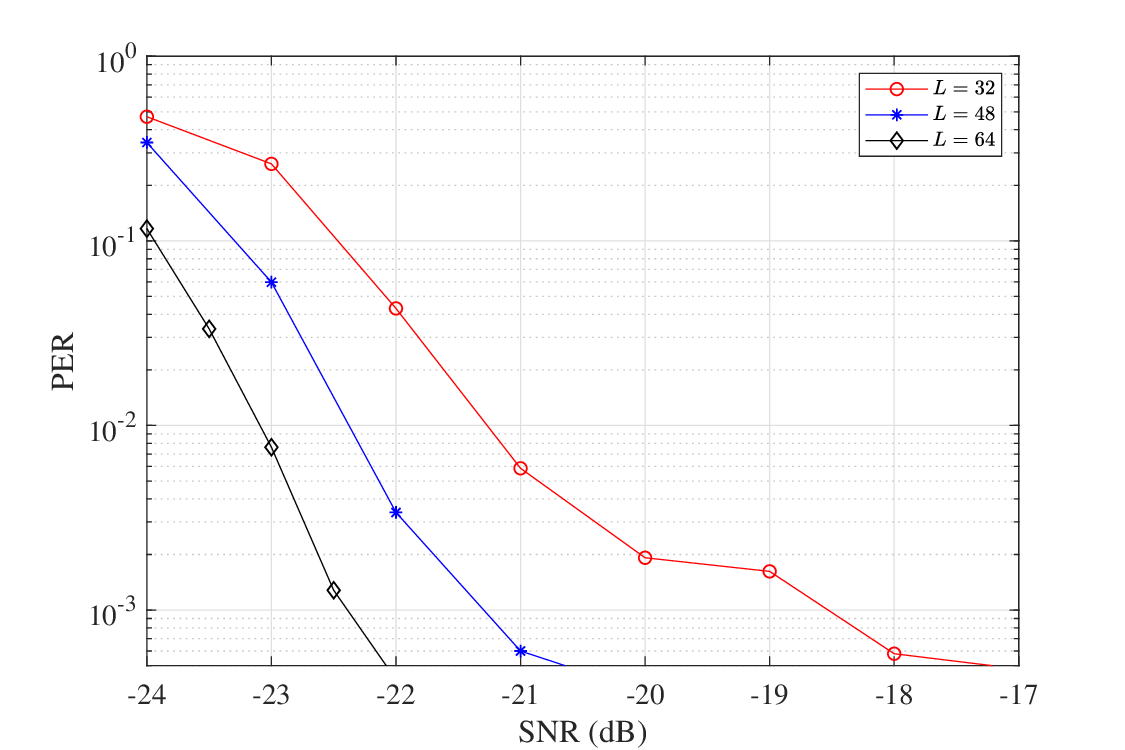}
\caption{PER with different numbers of antennas ($N=8$, 5 EDs)}.
%\label{fig.1}
\label{fig.PER2}
\end{figure}

Fig.~\ref{fig.PER2} shows the PER of the proposed preamble design with different numbers of antennas. With a higher number of antennas, the noise and interference can be further suppressed, which explains the declining PER when the number of antennas increases. It can be seen that by doubling the number of antennas from 32 to 64, the proposed design can save around 4 dB of transmit power to achieve the PER of $10^{-3}$.

\begin{figure}[t!]
\centering
\includegraphics[width=0.5\textwidth]{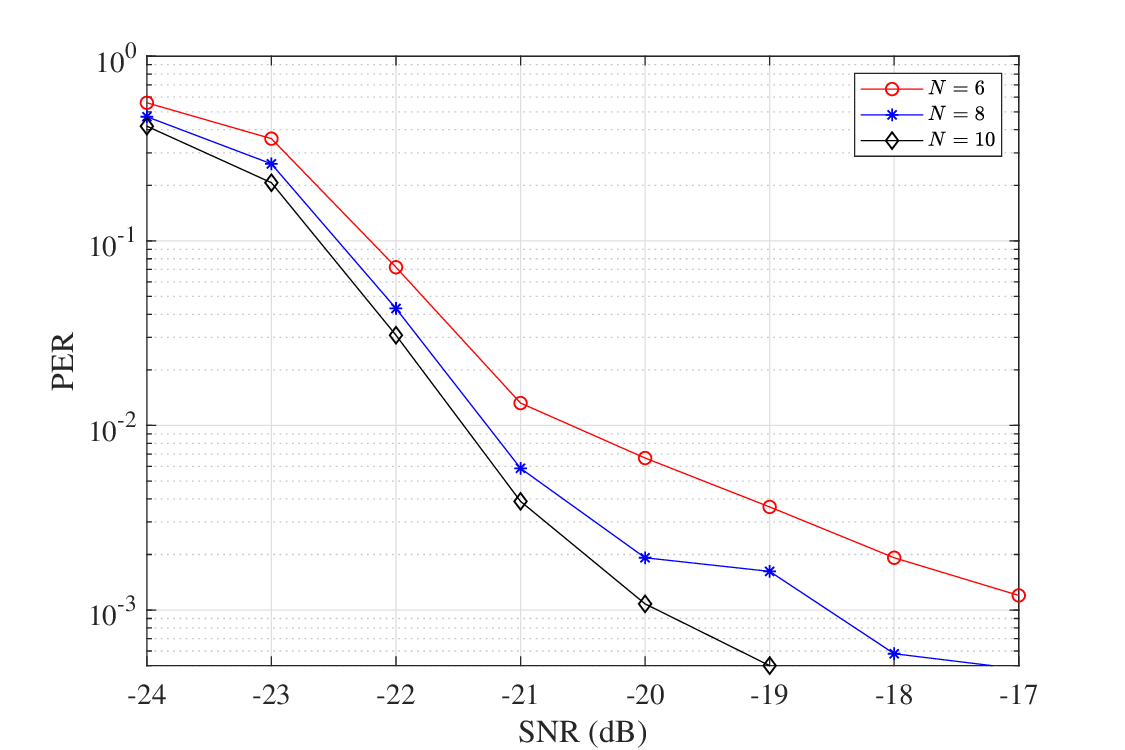}
\caption{PER with different preamble lengths ($L=32$, 5 EDs)}.
%\label{fig.1}
\label{fig.PER3}
\end{figure}
Fig.~\ref{fig.PER3} compares the PER performance with different preamble lengths. According to the standard \cite{semtech}, the length of the LoRa preamble is $N=6$ to 8 LoRa symbols. The difference of two symbols in preamble length results in a gap of $1.5$ dB to obtain the PER of $10^{-3}$. Extending the length of the preamble to 10 symbols leads to an extra gain of nearly 2 dB  in power to achieve the same PER. Although a longer preamble means more airtime to transmit a packet of the same payload, the power gain can still be a good trade-off, especially in IoT applications that require long battery life.

\section{Conclusion}\label{Sec:con}
This paper { proposed} a novel \emph{double-chirp} preamble design for multiuser massive MIMO LoRa networks, which allows LoRa GWs to detect asynchronous packets from multiple EDs simultaneously. First, we show that the challenge that prevents conventional \emph{single-chirp} preamble design from multiuser detection is the so-called \emph{preamble resemblance} effect. Then, we proposed the novel \emph{double-chirp} preamble design  to mitigate \emph{preamble resemblance}. Next, we proposed a ML-based preamble detection algorithm to increase the reliability and reduce confusion between true preambles and noise plus interference.  Finally, since our proposed method requires computation for every sample, we proposed a complexity reduction technique to efficiently calculate the DFT recursively rather than performing FFT every sampling period. Simulation results show that the proposed preamble design enables the GW to simultaneously detect preambles of multiple EDs with a reasonable power cost of roughly $2$ dB to increase the number of EDs from one to 15. It has also been shown that the use of multiple-antenna GW is necessary to mitigate the influence of noise and interference, which is evident from a gain of 4 dB achieved by doubling the number of antennas from 32 to 64.

\appendices

\section{Proof of Lemma \ref{lem1}} \label{appx-1}
According to \eqref{eq-pdw-Z-noise}, ${Z}_{n,u}$ is the sum of $L$ i.i.d. random variables. Thus, as a result of the central limit theorem \cite{Billingsley2005}, ${Z}_{n,u}$ can be approximated as a Gaussian random variable with the mean and variance being the sum of the means and variances of the addend random variables. Since $\tilde{W}_{n,\ell}[k]\sim\mathcal{CN}(0,\sigma^2)$, $\forall k$, it can be easily seen that:
%\begin{equation}
$\mathbb{E}\left\{\mathfrak{R}\left\{\left(\tilde{W}_{n,\ell}[\kappa_{u,1}]\right)^*\tilde{W}_{n,\ell}[\kappa_{u,2}]\right\}\right\}=0,$
%\end{equation}
and 
%\begin{equation}
$\text{Var}\left\{\mathfrak{R}\left\{\left(\tilde{W}_{n,\ell}[\kappa_{u,1}]\right)^* \tilde{W}_{n,\ell}[\kappa_{u,2}]\right\}\right\}=\frac{\sigma^4}{2}.$
%\end{equation}
Consequently, by adding up over all $L$ antennas, the approximated distribution of ${Z}_{n,u}$ in Lemma \ref{lem1} can be obtained. $\hfill\blacksquare$

\section{Proof of Lemma \ref{lem2}} \label{appx-2}
The proof of this lemma can be done similarly to the proof of Lemma \ref{lem1}, where the mean and variance of each individual addend random variable on the $\ell$th antenna is derived. From \eqref{eq-pdw-Z-PP}, it can be seen that when ${Z}_{n,u}$ is a PP:
%\begin{equation}
\begin{align}
\label{eq-pdw-Z_PProf}
&{Z}_{n,u}
%= \frac{1}{L}\mathfrak{R}\left\{ \check{\boldsymbol{V}}_{n,\kappa_{u,1}}^{\Hi}\check{\boldsymbol{V}}_{n,\kappa_{u,2}}\right\}\\
%=\frac{M}{2L}\Vert\boldsymbol{h}_u\Vert^2+\nn\\
%&   %\frac{1}{L}\mathfrak{R}\left\{\sqrt{\frac{M}{2}}\boldsymbol{h}_u^{\Hi}\left(\check{\boldsymbol{W}}_{n,\kappa_{u,1}}+\check{\boldsymbol{W}}_{n,\kappa_{u,2}}\right)+
%\check{\boldsymbol{W}}_{n,\kappa_{u,1}}^{\Hi}\check{\boldsymbol{W}}_{n,\kappa_{u,2}}\right\}\nn\\
%&+\\
=  \frac{1}{L}\sum_{\ell=1}^L\mathfrak{R}\left\{\sqrt{\frac{M}{2}}{h}_{u,\ell}^{*}\left(\tilde{{W}}_{n,\kappa_{u,1}}+\tilde{{W}}_{n,\kappa_{u,1}}\right)\right\}\nn\\
&+\frac{1}{L}\sum_{\ell=1}^L\mathfrak{R}\left\{	\left(\tilde{W}_{n,\ell}[\kappa_{u,1}]\right)^*\tilde{W}_{n,\ell}[\kappa_{u,2}]\right\}+\frac{M}{2L}\Vert\boldsymbol{h}_u\Vert^2.
%\xrightarrow[M\rightarrow\infty]{\mathrm{a.s}}\frac{M}{2},\nn
%\xrightarrow[M\rightarrow\infty]{\mathrm{a.s}}\frac{M}{2},\nn
\end{align}
%\end{equation}
The mean and variance of the three terms in \eqref{eq-pdw-Z_PProf} must be computed to derive the approximation of ${Z}_{n,u}$ as a PP.

\begin{itemize}
\item First, since ${h}_{u,\ell}$, $\tilde{{W}}_{n,\kappa_{u,1}}$ and $\tilde{{W}}_{n,\kappa_{u,2}}$ are independent, it can be easily seen that:
%%%%%%%%%%%%%%%%
%\begin{equation}\label{eq42}
$\mathbb{E}\left\{\mathfrak{R}\left\{{h}_{u,\ell}^{*}\left(\tilde{{W}}_{n,\kappa_{u,1}}+\tilde{{W}}_{n,\kappa_{u,1}}\right)\right\}\right\} = 0$,
%\end{equation}
%%%%%%%%%%%%%%%%%%%
and
%\begin{equation}\label{eq43}
%\begin{split}
$\mathrm{Var}\left\{\mathfrak{R}\left\{{h}_{u,\ell}^{*}\left(\tilde{{W}}_{n,\kappa_{u,1}}+\tilde{{W}}_{n,\kappa_{u,1}}\right)\right\}\right\} = \sigma^2$.
%\end{split}
%\end{equation}

\item The mean and variance of the second term have been derived in Lemma \ref{lem1}.

\item  Since $\boldsymbol{h}_{u,\ell}\sim\mathcal{CN}(0,1)$, we have
%\begin{equation}
$	\Vert\boldsymbol{h}_u\Vert^2\propto\chi^2\left(2L\right).$
%\end{equation}
Thus, 
$
\mathbb{E}\left\{\Vert\boldsymbol{h}_u\Vert^2\right\} = L
$
%%%%%%%%
and 
$
\mathrm{Var}\left\{	\Vert\boldsymbol{h}_u\Vert^2\right\} = L.
$
%%%%%%%%%%%%%%

\end{itemize}
Combining  all the means and variances above, the approximated Gaussian distribution of ${Z}_{n,u}$ as a PP can be obtained as in Lemma \ref{lem2} according to the central limit theorem \cite{Billingsley2005}. $\hfill\blacksquare$		

\section{Proof of Corollary \ref{col2}}\label{appx-3}
This proof shows how the pdf of $\boldsymbol{Z}_{n,u}$ can be derived when its first $p$ elements are PPs. The proof for  the  case when the last $p$ elements of  $\boldsymbol{Z}_{n,u}$ are PPs can be done similarly. 

First, in terms of the mean vector $\boldsymbol{\mu}_p$ of $\boldsymbol{Z}_{n,u}$, it can be easily seen that when the first $p$ elements are PPs, the mean value is $M/2$ according to Lemma \ref{lem2}, which means that $\mu_{p,i}=M/2$ for $i\leq p$. Whereas,  the remaining elements, which are noise, have zero mean.

In terms of the covariance matrix $\boldsymbol{\Sigma}_p$, first, the elements on the main diagonal is the variance of each individual element of $\boldsymbol{Z}_{n,u}$. Thus, for the first $p$ elements on the main diagonal of $\boldsymbol{\Sigma}_p$, which correspond to the $p$ first PPs, ${\Sigma}_{p,i,i}=\sigma_{\mathrm{PP}}^2$ for $i\leq p$. The remaining elements on the main diagonal $\boldsymbol{\Sigma}_p$ are the variance of noise, which is ${\Sigma}_{p,i,i}=\sigma_{\mathrm{n}}^2$ for $i> p$.

The elements on the anti-diagonal of $\mathbf{\Sigma}_p$ are the covariance between two different elements of $\boldsymbol{Z}_{n,u}$, without  loss of generality, say $Z_{n,u}$ and $Z_{n-M,u}$. There are three cases:
\begin{itemize}
\item If $Z_{n,u}$ and $Z_{n-M,u}$ are two PPs, they are correlated due to the common term ${M}\Vert\boldsymbol{h}_u\Vert^2/{2L}$ as can be seen from \eqref{eq-pdw-Z_PProf}, while other terms are uncorrelated. Thus, it can be  found that $
\mathrm{cov}\left\{Z_{n,u},Z_{n-M,u}\right\} = \mathbb{E}\left\{\frac{M	\Vert\boldsymbol{h}_u\Vert^2}{2L}\right\} =\frac{M^2}{4L}.
$
\item If either one of $Z_{n,u}$ and $Z_{n-M,u}$ is a PP, while the other is noise, they are uncorrelated, and hence,
$\mathrm{cov}\left\{Z_{n,u},Z_{n-M,u}\right\} = 0$.
\item If both $Z_{n,u}$ and $Z_{n-M,u}$ are noise, they are uncorrelated, and hence,
$\mathrm{cov}\left\{Z_{n,u},Z_{n-M,u}\right\} = 0$.
\end{itemize}
From these  cases, the elements on the anti diagonal of the covariance matrix $\boldsymbol{\Sigma}_p$ can be obtained as in Corollary \ref{col2}. $\hfill\blacksquare$

\bibliographystyle{IEEEtran}

\balance
\end{document}